\documentclass[aps,prl,reprint,amsmath,amssymb,notitlepage,citeautoscript,superscriptaddress]{revtex4-1}

\usepackage{bm,mathrsfs,dcolumn,graphicx,color}
\usepackage{amsmath}
\usepackage{graphicx}
\usepackage[T1]{fontenc}
\usepackage{hyphenat}
\usepackage{setspace}

\usepackage{amssymb} 
\usepackage{wasysym} 
\usepackage{siunitx} 

\usepackage[normalem]{ulem}
\definecolor{darkerblue}{rgb}{0,0,0.75}
\definecolor{darkerred}{rgb}{0.8,0,0}
\usepackage[colorlinks=true,citecolor=darkerblue,linkcolor=darkerred]{hyperref}

\newcommand{\supp}{See Supplemental Material at [url], which includes Refs.~[..]--[..], for additional experimental details and theoretical analysis.}

\definecolor{ablue}{rgb}{0.1,0.3,0.65}

\begin{document}

\title{Autoionization and dressing of excited excitons by free carriers in monolayer WSe$_2$}






\author{Koloman Wagner}
\author{Edith Wietek}
\author{Jonas D. Ziegler}
\affiliation{Department of Physics, University of Regensburg, Regensburg D-93053, Germany}
\author{Marina Semina}
\affiliation{Ioffe Institute, Saint Petersburg, Russian Federation}
\author{Takashi Taniguchi}
\affiliation{International Center for Materials Nanoarchitectonics,  National Institute for Materials Science, Tsukuba, Ibaraki 305-004, Japan}
\author{Kenji Watanabe}
\affiliation{Research Center for Functional Materials, National Institute for Materials Science, Tsukuba, Ibaraki 305-004, Japan}
\author{Jonas Zipfel}
\affiliation{Department of Physics, University of Regensburg, Regensburg D-93053, Germany}
\author{Mikhail M. Glazov}
\affiliation{Ioffe Institute, Saint Petersburg, Russian Federation}
\author{Alexey Chernikov\footnote{alexey.chernikov@ur.de}}
\affiliation{Department of Physics, University of Regensburg, Regensburg D-93053, Germany}

\begin{abstract}
We experimentally demonstrate dressing of the \emph{excited} exciton states by a continuously tunable Fermi sea of free charge carriers in a monolayer semiconductor.
It represents an unusual scenario of \emph{two-particle excitations} of charged excitons previously inaccessible in conventional material systems.
We identify excited state trions, accurately determine their binding energies in the zero-density limit for both electron- and hole-doped regimes, and observe emerging many-body phenomena at elevated doping.
Combining experiment and theory we gain access to the intra-exciton coupling facilitated by the interaction with free charge carriers. 
We provide evidence for a process of \emph{autoionization} for quasiparticles, a unique scattering pathway available for excited states in atomic systems.
Finally, we demonstrate a complete transfer of the optical transition strength from the excited excitons to dressed excitons, Fermi polarons, as well as the associated light emission from their non-equilibrium populations.
\end{abstract}
\maketitle

Interactions between electronic quasiparticles are of fundamental interest in solid state physics.
A particularly intriguing scenario is the coexistence of tightly-bound electron-hole states, known as excitons, with a Fermi sea of free carriers.
This many-body problem leads to a rich variety of phenomena, including bound three-particle states, or trions\,\cite{Lampert1958,Kheng1993,PhysRevB.53.R1709,Astakhov1999,Sergeev2001,Mak2012,Ross2013,Ganchev2015}, exciton-electron interactions in the Fermi polaron picture\,\cite{Koudinov2014,Sidler2016,Efimkin2017,Chang2018,Fey2020,Cotlet2020}, renormalization effects from screening and Pauli blocking\,\cite{Haug1989} as well as high-density phases beyond the Mott-transition\,\cite{Klingshirn2007}.
Their physics is traditionally explored in low-dimensional systems, such as semiconductor quantum wells and, more recently, layers of transition-metal dichalcogenides (TMDCs)\,\cite{Mak2012,Ross2013,Mouri2013,Berkelbach2013,Chernikov2015b,Singh2016,Plechinger2016,Courtade2017,Vaclavkova2018,VanTuan2019,Carbone2020}.
The latter offer a highly suitable platform with strong Coulomb interaction and excellent tunability\,\cite{Xu2014,Yu2015,Xiao2017,Wang2018}.

\begin{figure}[t]
	\centering
			\includegraphics[width=8.65 cm]{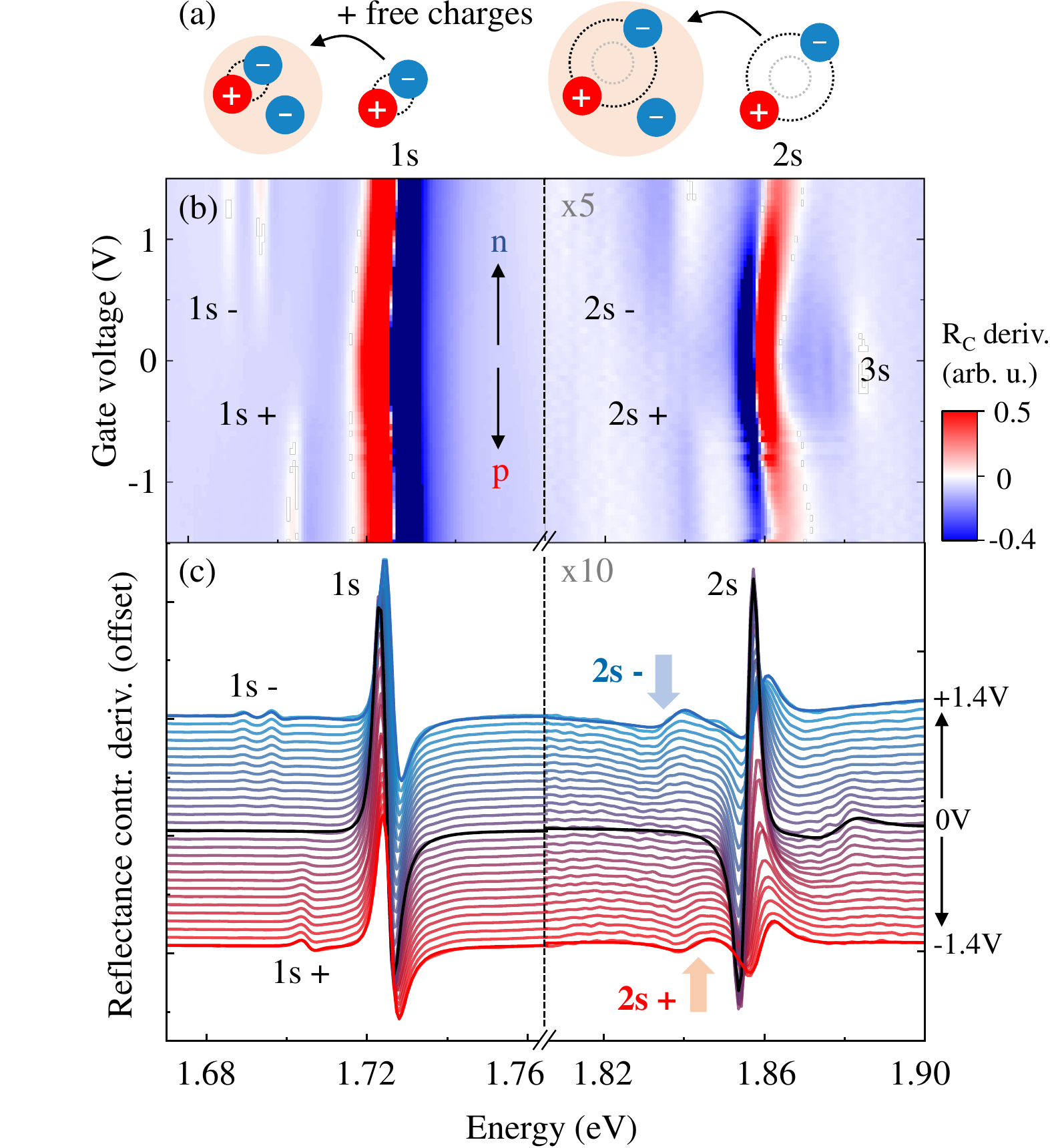}
		\caption{(a) Schematic illustration of ground and excited state excitons in the presence of free charges. 
		(b) Reflectance contrast derivative spectra from an electrically-tunable WSe$_2$ monolayer in the low-density regime. 
		The gate voltage of 1\,V corresponds to an estimated free carrier density on the order of $10^{11}$\,cm$^{-2}$.
		(c) Corresponding selected spectra, vertically offset in steps of 0.1\,V. 
		Simulated curves are shown in bold for -1.4, 0, and 1.4\,V.
		}
	\label{fig1}
\end{figure}

So far, however, the coupling of excitons to free charge carriers has been studied exclusively for the exciton \emph{ground} state.
For quantum wells, the access to the \emph{excited} states under electrical doping was limited by weak Coulomb interactions.
For TMDCs, early interpretations of electrical tunability\,\cite{Chernikov2015b} were severely affected by environmental inhomogeneities\,\cite{Raja2019,Rhodes2019}, dominating excited state properties.
Only recently, clean systems, free from long-range disorder became available\,\cite{Dean2010,Cadiz2017, Ajayi2017, Courtade2017, Manca2017, Stier2018}.
These now offer a direct access to a highly unusual scenario for quasiparticles in condensed matter that involves \emph{excited} exciton states dressed by the coupling to the Fermi sea of free charge carriers, as illustrated in Fig.\,\ref{fig1}\,(a).
Conceptually reminiscent of \emph{two-electron excitations} in H$^-$\,\cite{Rau1996}, these complexes allow for studies of interacting bound and free carrier mixtures in metastable, excited states\,\cite{Yan2020}.
First indication of these phenomena was recently reported in Ref.~\cite{Arora2019} for an unintentionally doped WS$_2$ monolayer without carrier density control\,\cite{noteA}.

In our joint experiment-theory study we provide extensive evidence for 
dressing of \emph{excited excitons} by the interactions with the Fermi sea of free charge carriers. 
Field-effect transistors based on hBN-encapsulated WSe$_2$ are employed for continuous electrical tuning of free carrier densities during optical measurements.
The approach enables unambiguous identification of the excited state trions, clear determination of their binding energies in both electron- and hole-doped regimes, as well as the demonstration of the \emph{autoionization}.
The latter is a highly efficient process known from atomic physics that is unique for two-electron excitations of ions\,\cite{Rau1996} and is consistent with the discussed theoretical properties of the excited state trions.
At elevated densities, we observe a \emph{complete transfer} of the oscillator strength to the excited exciton states dressed by the Fermi sea.
Finally, we demonstrate that these many-body complexes can host transient populations and emit light that emerges as non-equilibrium, hot luminescence.

The studied devices were fabricated by mechanical exfoliation and dry-stamping\,\cite{Castellanos-Gomez2014} of bulk crystals on pre-patterned gold electrodes.
WSe$_2$ monolayers were encapsulated between thin hBN layers and we used few-layer graphene as top- and bottom-gates as well as for source-drain contacts.
The devices were placed in a microscopy cryostat and cooled down to a nominal heat-sink temperature of 5\,K.
In optical measurements we used a spectrally broad tungsten-halogen lamp for reflectance and a 532\,nm continuous-wave laser for photoluminescence (PL).
The incident light was focused onto the sample by a glass-corrected 60x microscope objective.
The reflected and emitted signals were dispersed in a spectrometer and detected by a charge-coupled-device camera. 
Additional details on sample preparation and experimental procedures are given in the Supplemental Material (SM)\,\footnote{\supp}.

Representative reflectance contrast derivatives are presented in Fig.\,\ref{fig1} as a function of gate voltage in a color plot in (b) and as individual spectra in (c).
The reflectance contrast is defined as $(R_{s}-R_{ref})/R_{ref}$, where $R_{s}$ and $R_{ref}$ are reflectance spectra with and without the WSe$_2$ monolayer, respectively.
The presented range of gate-voltages corresponds to the low-density region with free carrier densities up to 10$^{11}$\,cm$^{-2}$. 
The response at neutrality point is dominated by the exciton ground state ($1s$) at 1.72\,eV with the first and second excited states ($2s$ and $3s$) located at 1.85 and 1.88\,eV, respectively.
At finite gate-voltages additional resonances appear on the low-energy side of $1s$.
These are labeled by the sign of the free carriers charge ($+/-$) and are commonly understood either as ground-state trions\,\cite{Mak2012,Ross2013} or attractive Fermi polarons\,\cite{Ganchev2015,Sidler2016,Efimkin2017}. 
The two descriptions are essentially equivalent at low charge densities\,\cite{Glazov2020a}.

\begin{figure}[t]
	\centering
			\includegraphics[width=8.5 cm]{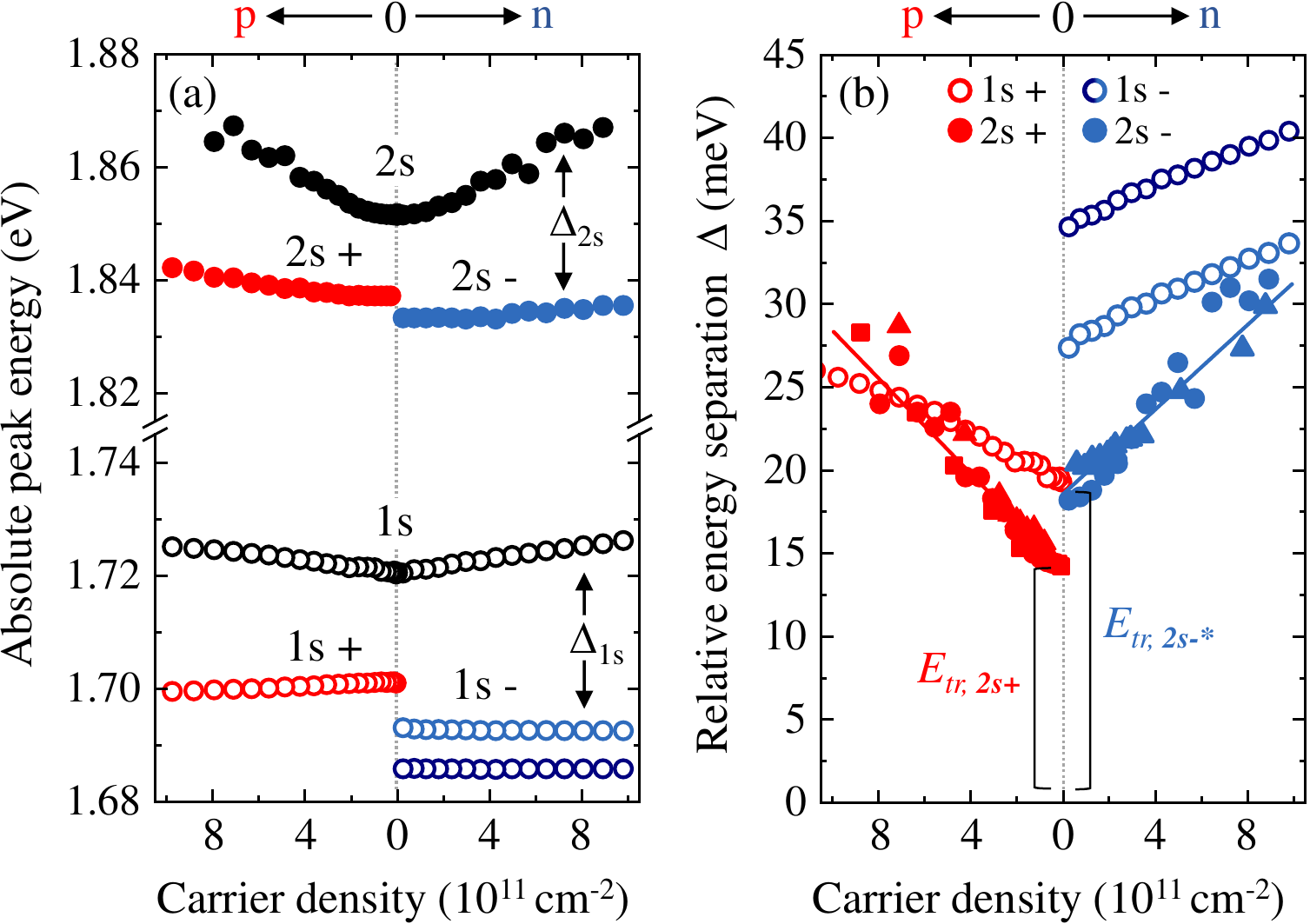}
		\caption{(a) Absolute peak energies for the ground and first excited state resonances as function of the free electron and hole densities. 
		(b) Corresponding relative energy separation $\Delta$ obtained by subtracting the charged exciton energy from that of the respective neutral state. 
		The $2s\pm$ data ($\fullmoon$,$\triangle$,$\square$) is aggregated from several devices and sample positions (see SM for additional overview).
		}
	\label{fig2}
\end{figure}

Interestingly, very similar features emerge in the spectral region of the $2s$ resonance on both electron- and hole-doped sides. 
Their energies are summarized in Fig.\,\ref{fig2}\,(a) as function of the free carrier densities in a broader doping range, where additional many-body renormalization effects appear.
The peak parameters are extracted from fitting the reflectance contrast spectra with model dielectric functions using a transfer-matrix approach\,\cite{Byrnes2016,Raja2019,Brem2019}.
The density is estimated from the linear scaling of the energy separation $\Delta_{1s\pm}$ between $1s$ and $1s\pm$ states (see SM).
While the relative energy shifts are very similar for the $2s\pm$, as illustrated in Fig.\,\ref{fig2}\,(b), they exhibit notable differences.
First, the zero-density limit of $\Delta_{2s\pm}$ is lower in contrast to $\Delta_{1s\pm}$.
This is consistent with the identification of the $2s\pm$ states as excited state trions (attractive Fermi polarons), since these values represent trion binding energies of $E_{tr,2s+}=14.1$\,meV and $E_{tr,2s-^*}=18.6$\,meV that are expected to be smaller for an excited state (c.f. 19.4\,meV for $1s+$, 27.9 and 34.7\,meV for $1s-$ doublet).
Here, the binding energy of $2s-$ is an average value.
Due to a larger broadening we cannot exclude the presence of a doublet (related to the valley fine structure) that seems weakly indicated in the spectra.

Secondly, the increase of the energy separation $\Delta$ is steeper for $2s\pm$ by almost a factor of two.
We note, that both in the trion and polaron descriptions $\Delta$ is proportional to the Fermi-energy of the free charges due to momentum-conservation\,\cite{Esser2001} and exciton-electron interactions\,\cite{Sidler2016}, respectively. 
In the former approximation, however, one would generally expect the slope to be exactly the same for all trion states.
Most importantly, the $2s\pm$ trion binding energies are roughly on the order of the \emph{total} binding energy of the $2s$ exciton\,\cite{Raja2019,Goryca2019}. 
It is in stark contrast to the conventional picture of the ground state trions that exhibit rather weak binding of an additional charge carrier to a tightly-bound electron-hole core.
This does not hold for the excited states and is likely related to the observed changes in the interactions.

It is thus instructive to consider the coupling of the excitons to free charge carriers in a general theoretical framework.
In atoms, the $2s\pm$ states would correspond to \emph{two-electron} excitations of a {H$^-$} ion\,\cite{Rau1996}, where both electrons are excited to states resembling higher-lying orbitals with finite binding energies in contrast to one-electron excitations\,\cite{Hill1977}.
Here, the key differences are (i) comparable values of the masses and (ii) presence of the finite densities of resident carriers.
We extend the Fermi-polaron formalism\,\cite{Suris2001,Sidler2016,Efimkin2017,Cotlet2019,Glazov2020a} to include the excited states describing exciton-electron interaction via short-range potential (see SM for details).
Let 
\begin{align}
\label{greens:x}
G_j(\varepsilon, \bm k) = \frac{1}{\varepsilon - E_j - \frac{\hbar^2 k^2}{2m_x} + \mathrm i 0},
\end{align} 
be the exciton (retarded) bare Greens function found at the negligible damping, $E_j$ is its energy, and $m_x=m_e+m_h$ the exciton translational mass, i.e., the sum of the electron ($m_e$) and hole ($m_h$) masses. 
The subscript $j=1s,2s,\ldots$ enumerates relative motion states.
The equations for the scattering amplitudes $T_{ij}$ in the non-self-consistent approximation read
\begin{align}
\label{scattering:T}
&T_{ij}(\varepsilon,\bm k) = V_{ij} \\\notag
&+\sum_l V_{il} \sum_{\bm p}(1-n_{\bm p}) G_l\left(\varepsilon - \frac{\hbar^2p^2}{2m_e}, \bm k - \bm p\right) T_{lj}(\varepsilon,\bm k).
\end{align}
Here, we introduce the matrix elements $V_{ij}$ for the scattering from the exciton state $j$ to $i$ and assume that these are small, i.e., $\left|V_{ij}\right|\mathcal D \ll 1$, where $\mathcal D = \mu_{eX}/(2\pi \hbar^2)$ is the reduced electron-exciton density of states and $\mu_{eX}^{-1}= m_e^{-1} + m_x^{-1}$ is the electron-exciton reduced mass.
We further consider negligible Pauli blocking with the phase-filling factor $(1-n_{\bm p})\approx 1$, as well as the excitations within the light cone and total momentum $\bm k\approx 0$. 
To simplify the following calculations and obtain the analytical result we consider only the coupling between the ground and excited excitonic states, $1s$ and $2s$ (omitting $s$ in the notation).
The energy-dependent scattering amplitude $T_{22}\equiv T_{22}(\varepsilon,0)$ for $2s$ satisfies the equation
\begin{align}
\label{T22}
T_{22} &= (1+S_2 T_{22}) \left[V_{22} +\frac{|V_{12}|^2 S_1}{1-V_{11} S_1} \right]
\end{align}
with $S_l=\sum_{\bm p} G_{l}(\varepsilon,\bm p)$. In the vicinity of $\varepsilon = E_{2s}$, Eq.~\eqref{T22} has a pole in a complex energy plane representing the ``excited trion resonance'' at 
\begin{align}
\label{2s:trion}
\varepsilon &= E_{2s} - E_{tr,2s}\left(1+ \mathrm i \tan{\phi}\right),
\end{align}
with the excited trion binding energy (for $V_{22}<0$):
\begin{align}
\label{binding}
E_{tr,2s} &\approx  E_{2s} \exp{\left(\frac{1}{\mathcal D V_{22}} \right)}\cos{\phi}, 
\end{align}
and a phase $\phi$ from a finite $2s\to 1s$ coupling ($V_{12}$),
\begin{align}
\label{phi}
\phi \approx \pi  \frac{ |V_{12}|^2}{|V_{22}|^2}.
\end{align}
Here the approximate equalities are obtained in the leading order in $\mathcal D V_{ij}\ll 1$ and we note that $\phi$ should not exceed $\pi/2$ ($|V_{12}| \lesssim |V_{22}|$) as the binding energy becomes negative otherwise.
\begin{figure}[t]
	\centering
			\includegraphics[width=7.5 cm]{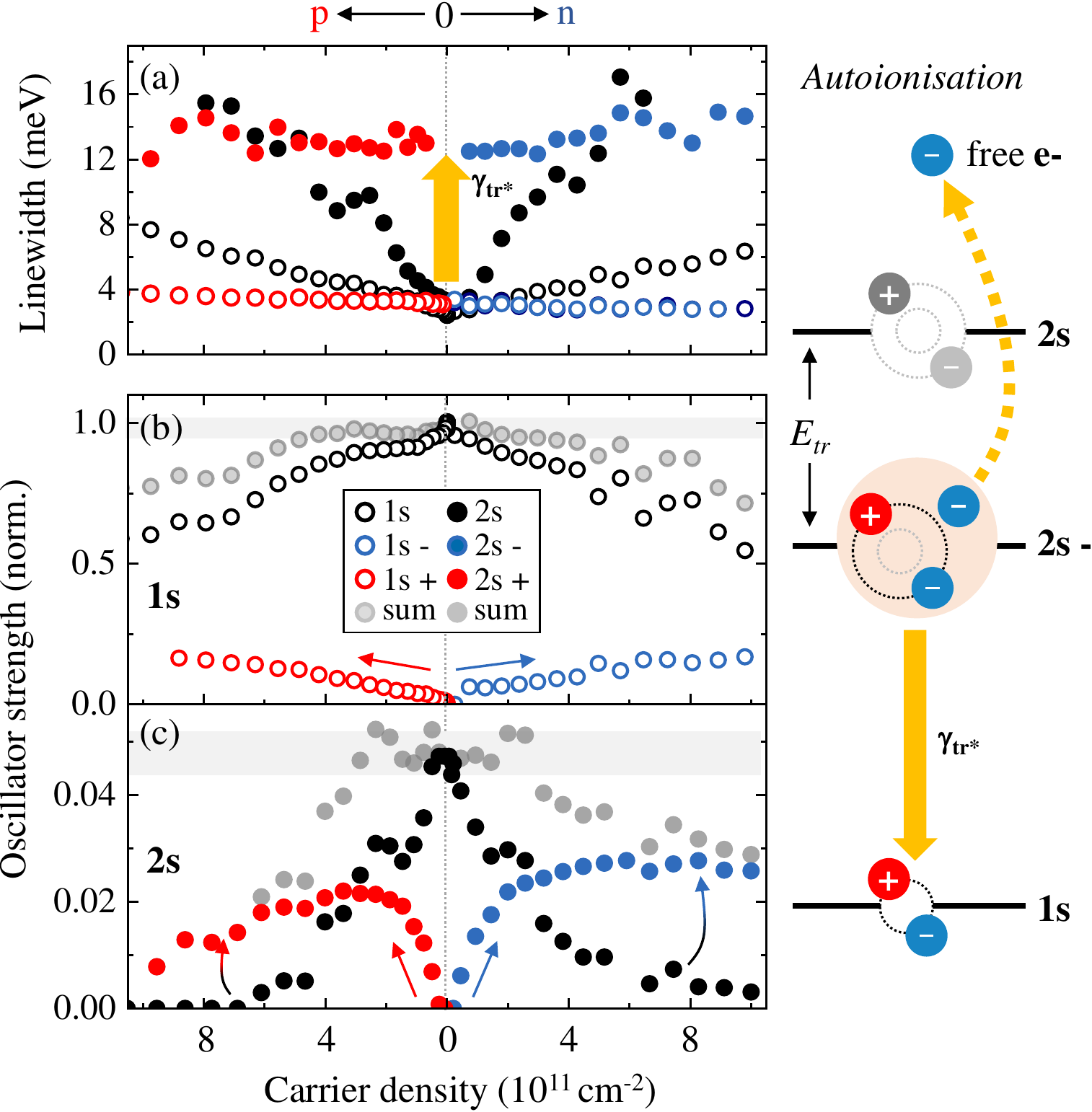}
		\caption{(a) Extracted non-radiative linewidths presented as full-width-at-half-maxima. 
			(b) and (c) Corresponding oscillator strengths, normalized to that of the exciton ground state ($1s$) at charge neutrality conditions.
		Gray areas indicate constant total oscillator strength.
		All figures share the same legend included in (b).
		Schematic illustration of the autoionization process is shown on the right.
		}
	\label{fig3}
\end{figure}

This result has two important consequences intimately associated with $2s\pm$ being an excited state.
Following Eqs.\,\eqref{binding} and \eqref{phi}, the trion binding energy is renormalized by the coupling $V_{12}$ between $2s$ and $1s$ (similarly to the presence of higher excited states, see SM). 
Depending on the value of $\phi$ which could be tuned, in principle, by tailoring $V_{ij}$ via the engineering of the dielectric environment, the bound state may or may not be present. 
Importantly, in contrast to the standard Fermi-polaron model, the resonance energy in Eq.\,\eqref{2s:trion} includes an \emph{imaginary} component, i.e., an intrinsic damping of $2s\pm$:
\begin{equation}
\label{gammaExc}
\gamma_{tr*} = E_{tr,2s} \tan{\phi}.
\end{equation}
It is a consequence of the specific process for two-electron excited states known as \emph{autoionization} in atomic physics:
\[
(2s +\mbox{electron})_{\,bound} \to 1s + \mbox{electron}'_{\,free}.
\]
Here, an excited state with an additional bound electron relaxes to the ground state, transferring the excess energy to that electron which becomes unbound, see schematics in Fig.\,\ref{fig3}.
In the context of quasiparticles in solids it is essentially the \emph{intra}-excitonic Auger effect. 

Autoionization should thus lead to a characteristic, finite broadening that is present exclusively for the charged excited states.
In addition, it should be initially independent from the charge carrier density as it stems entirely from the scattering processes within a single composite quasiparticle.
To test this prediction we extract the non-radiative linewidths from the reflectance contrast data, presented in Fig.\,\ref{fig3}\,(a).
Importantly, the linewidths of the $1s$, $1s\pm$, and $2s$ states are all very narrow, on the order of a few meV in the zero-doping limit.
None of them can be subject to autoionization.
In stark contrast to that, both $2s+$ and $2s-$ resonances exhibit an additional, density-independent broadening $\gamma_{tr*}$ as large as 10\,meV.
These combined results further exclude other alternatives that would either apply to all exciton states (e.g., inhomogeneous broadening) or be density-dependent (e.g., enhanced direct scattering with free electrons). 

Altogether, the observations of excited $2s\pm$ states with finite binding energies and characteristic broadening from autoionization are consistent with the theoretical description.
Further considering the $1s$ trion binding energy of $E_{tr,1s}=\bar E_{1s} \exp[{1/(\mathcal D V_{11})}]$ we can extract the interaction matrix elements $V_{ij}$ from the experimentally obtained $2s$ binding energies and linewidths using Eqs.\,\eqref{binding} and \eqref{gammaExc}.
For the hole-doped side we find $V_{11}$ and $V_{12}$ on the order of 100\,eV$\si{\angstrom}^{2}$ and $V_{22}$ of about 200\,eV$\si{\angstrom}^{2}$ (numbers for the electron side are very similar). 
Interestingly, the free-carrier-mediated $2s\to 1s$ scattering ($V_{12}$) is indeed very efficient, as reflected in the fast autoionization time on the order of 60\,fs from the 10\,meV broadening of the $2s\pm$.
It is further reasonable that $|V_{22}|>|V_{11}|$, since the interaction within the $2s$ state with free charges ($V_{22}$) should be stronger than for the $1s$ due to a larger spatial extension of the $2s$ wavefunction\,\cite{Stier2018,Zipfel2018,Goryca2019}. 
We note, however, that the obtained parameters also indicate the limits of the basic model with $\left|V_{ij}\right|\mathcal D$ approaching unity.

Closely related is the transfer of the oscillator strength between neutral and charged states as a function of free carrier density.
It is observed both for $1s$ and $2s$, as illustrated in Fig.\,\ref{fig3}\,(b) and (c), respectively.
The exchange occurs at much lower carrier densities for the excited state, however, reaching equal magnitudes of $2s$ and $2s\pm$ at about $3\times10^{11}$\,cm$^{-2}$ that is generally consistent with a larger radius and stronger interactions.
Interestingly, at higher doping towards roughly $8\times10^{11}$\,cm$^{-2}$, the oscillator strength is completely transferred to the $2s\pm$ states with that of the $2s$ reaching essentially zero. 
Again, the similarity of the $E_{2s}$ and $E_{2s,\pm}$ binding energies render perturbative description of the exciton -- Fermi-sea coupling inapplicable~\cite{Glazov2020a} and can be responsible for the observed effect.
Here, the $2s\pm$ states coexist with the ground state that is still largely dominated by the ``neutral'' $1s$ resonance, yet are far from the Mott-transition that should occur above $10^{12}$\,cm$^{-2}$.
These results may also have consequences for previous measurements of the excited states in non-encapsulated, unintentionally doped samples\,\cite{Chernikov2014, Chernikov2015b}, albeit both differences in binding energies and the presence of disorder should affect a more direct comparison.
\begin{figure}[t]
	\centering
			\includegraphics[width=8.4 cm]{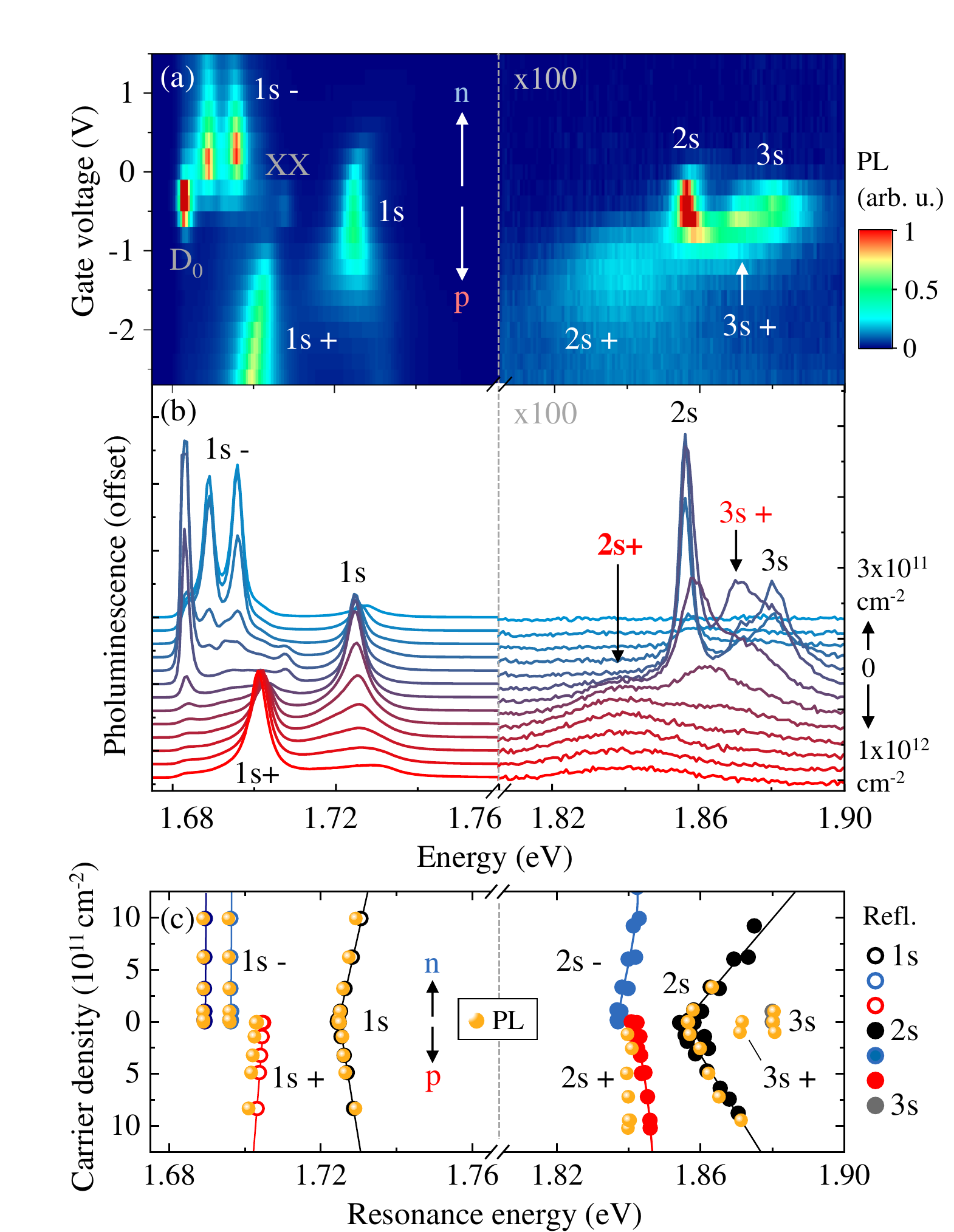}
		\caption{(a) Photoluminescence spectra as a function of gate voltage from continuous-wave excitation with a power density of 250\,W/cm$^2$ at the photon energy of 2.3\,eV.
		The gate voltage of 1\,V corresponds to an estimated free carrier density on the order of $10^{12}$\,cm$^{-2}$.
		Emission below the $D_0$ peak is suppressed by a spectral filter. 
		(b) Corresponding selected spectra, vertically offset with gate-voltage steps of 0.2\,V. 
		(c) Extracted peak energies from PL compared to those from reflectance.
		}
	\label{fig4}
\end{figure}

Finally, we demonstrate that electrically tunable mixing of excited states with the Fermi sea can host finite populations and be detected in emission-type measurements.
Continuous-wave PL data are presented in Fig.\,\ref{fig4} as color maps and as selected spectra in (a) and (b), respectively.
In the spectral range of $1s$ we observe the typical response from bright and spin-dark excitons as well as a weak biexciton emission at charge neutrality conditions.
With increasing gate voltage, $1s\pm$ states emerge\,\cite{Mak2012,Ross2013,Mouri2013,Plechinger2016,Courtade2017,Vaclavkova2018,VanTuan2019} and the $1s$ PL is increasingly quenched. 
Similar features appear at the low energy side of the $2s$ and $3s$ transitions in the hole-doped region corresponding to the $2s+$ state and even an indication of a $3s+$.
Considering the nominal lattice temperature of 5\,K, the emission from excited states should occur far from the equilibrium and appear during relaxation after off-resonant injection.
At the electron-doped side, however, the luminescence is strongly suppressed both for the ground and excited states.
Altogether, the PL closely follows the reflectance, as illustrated by summarizing the peak energies obtained by the two methods in Fig.\,\ref{fig4}\,(c).
Only at elevated densities there are indications of a finite, density-dependent Stokes shift for $2s+$\,\cite{noteA} that may be connected to the effects recently proposed in Ref.\,\onlinecite{Cotlet2020}.

In summary, we have explored a highly unusual scenario of \emph{excited} exciton states being dressed by the interactions with a continuously-tunable Fermi sea of free charges under both electron- and hole-doping conditions.
In the low-density regime these states are conceptually similar to the two-electron excitations known from the negatively charged hydrogen ion.
As demonstrated by the experiment and theory they exhibit characteristic properties including a process known as autoionization from atomic physics.
Autoionization becomes uniquely possible for double-excited states and intrinsically limits their lifetime to about 60\,fs in the studied WSe$_2$ monolayers.
In contrast to the ground state, excited states are subject to distinctly different many-body renormalization of their energies at increased doping levels as well as a more efficient transfer of the oscillator strength up to a complete exchange due to a larger size.
Finally, despite intrinsically short lifetimes, such states can be transiently populated by optical pumping and emit light with a very high sensitivity to electrical gating.
Overall, these results open up avenues to explore intricate, previously inaccessible many-body physics of quasiparticles in condensed matter to foster a deeper understanding of strong electronic interactions in low-dimensional materials.

\section{Acknowledgments}
We thank Viola Zeller and Hedwig Werner from the group of Dominique Bougeard for their support with the sample preparation.
Financial support by the DFG via Emmy Noether Initiative (CH 1672/1-1) and SFB 1277 (project B05) is gratefully acknowledged.
M.A.S. is gratefully acknowledging financial support from Russian Science Foundation (project no. 19-12-00273).
K.W. and T.T. acknowledge support from the Elemental Strategy Initiative, conducted by the MEXT, Japan, Grant Number JPMXP0112101001, JSPS KAKENHI Grant Numbers JP20H00354 and the CREST (JPMJCR15F3), JST.




%

\end{document}


\title{Supplemental Material:\\ Autoionization and dressing of excited excitons by free carriers in monolayer WSe$_2$}

\author{Koloman Wagner}
\author{Edith Wietek}
\author{Jonas D. Ziegler}
\affiliation{Department of Physics, University of Regensburg, Regensburg D-93053, Germany}
\author{Marina Semina}
\affiliation{Ioffe Institute, Saint Petersburg, Russian Federation}
\author{Takashi Taniguchi}
\affiliation{International Center for Materials Nanoarchitectonics,  National Institute for Materials Science, Tsukuba, Ibaraki 305-004, Japan}
\author{Kenji Watanabe}
\affiliation{Research Center for Functional Materials, National Institute for Materials Science, Tsukuba, Ibaraki 305-004, Japan}
\author{Jonas Zipfel}
\affiliation{Department of Physics, University of Regensburg, Regensburg D-93053, Germany}
\author{Mikhail M. Glazov}
\affiliation{Ioffe Institute, Saint Petersburg, Russian Federation}
\author{Alexey Chernikov\footnote{alexey.chernikov@ur.de}}
\affiliation{Department of Physics, University of Regensburg, Regensburg D-93053, Germany}

\maketitle

{\hypersetup{hidelinks}
\tableofcontents
}

\newpage

\section{Device fabrication}

\begin{figure}[h]
	\centering
			\includegraphics[width=13 cm]{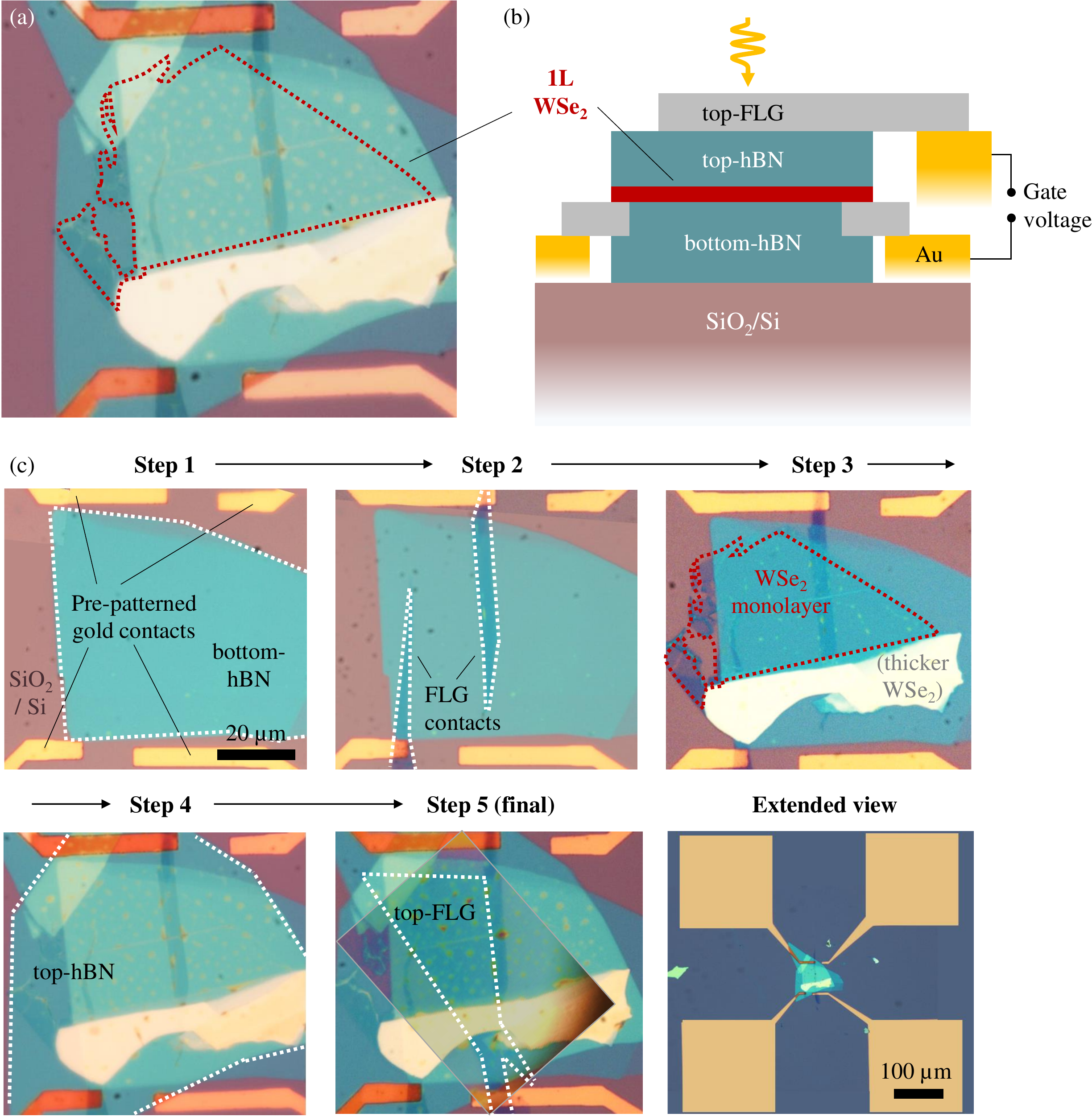}
		\caption{(a) Optical micrograph of a gate-tunable field-effect transistor device based on a hBN-encapsulated WSe$_2$ monolayer. 
			(b) Schematic representation of the corresponding structure.
			(c) Series of micrographs illustrating individual stacking steps (an additional image with higher contrast is included for step 5).
		}
	\label{figS1}
\end{figure}

All optical measurements in our study are performed on gate-tunable tungsten diselenide monolayers (1L WSe$_2$) encapsulated in high-quality hexagonal boron nitride (hBN) with 5 to 10\,nm thickness.
The hBN layers reduce environmental disorder effects and act as dielectric barriers for electrical gating.
Thin few-layer graphene flakes with a thickness of a few to 10\,nm are used as top- and bottom-gates as well as source- and drain-contacts to the WSe$_2$ layer.
A representative optical micrograph of one of the several fabricated devices used in our study is presented in Fig.\,\ref{figS1}\,(a).
A schematic illustration of a generic device structure is shown in Fig.\,\ref{figS1}\,(b).

The devices are fabricated by micro-mechanical exfoliation of bulk crystals and subsequent all-dry viscoelastic stamping\,\cite{Castellanos-Gomez2014} in the following manner, see Fig.\,\ref{figS1}\,(c).
We use commercially available WSe$_2$ and graphite crystals from ``HQGraphene'' and hBN from NIMS.
Bulk material is mechanically cleaved and exfoliated onto PDMS films. 
After identifying suitable layers by optical contrast, a thin hBN is stamped on top of a preheated SiO$_2$/Si substrate with pre-patterned gold contacts.
Two elongated few-layer graphene (FLG) flakes are placed on top of the hBN, connecting it to the two gold contacts, and a WSe$_2$ monolayer is subsequently stamped on top. 
Then, the second hBN flake is placed onto the remaining bare area of the WSe$_2$ layer and the device is completed by stamping a larger few-layer graphene flake on top of the stack. 
The latter acts as a top-gate, connecting it to another gold contact. 
After each step the stack is annealed in high vacuum at 150\,$^\circ$C for one to three hours. 
In one of the studied devices the structure is extended by an additional thin graphite flake placed on top of the SiO$_2$/Si substrate prior to stamping the remaining device structure and is used as a bottom-gate.

\newpage
\section{Electrically-tunable reflectance contrast}
\subsection{As-measured reflectance contrast and first derivative spectra}

\begin{figure}[hb]
	\centering
			\includegraphics[width=15.5 cm]{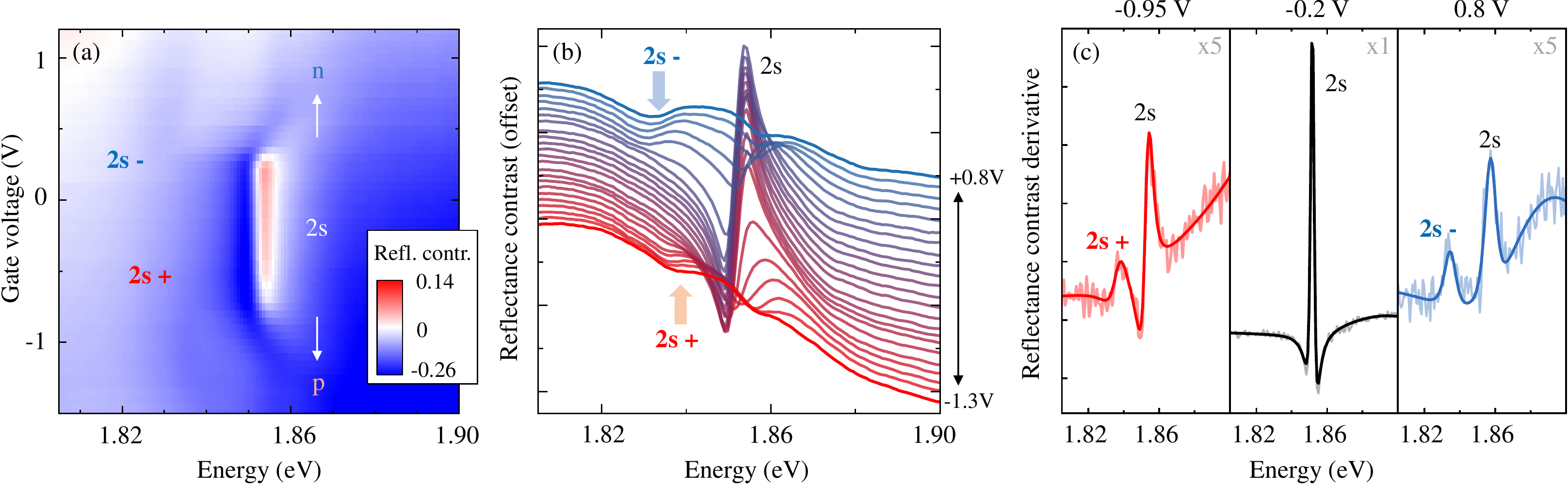}
		\caption{(a) As-measured reflectance contrast spectra of the WSe$_2$ monolayer (device A, position II) in the range of excited state excitons as a function of applied gate voltage (area without the monolayer is used as a reference).
				First excited state is labeled by $2s$ in a hydrogen-like notation and ($+/-$) indicates the sign of the free carriers charge.
		(b) Corresponding selected spectra, vertically offset for clarity. 
		The overall spectral shape is determined by the optical interference effects in the studied multi-layer structure.
		Small, gate voltage-dependent changes in the overall tilt are likely due to minor spatial drifts and changes in focus during longer measurements. 
		(c) First derivative spectra, taken directly from the reflectance contrast data in $p$-doped ($-0.95$\,V), neutral ($-0.2$\,V), and $n$-doped (+0.8\,V) regimes.
		Simulated curves obtained from multi-Lorentzian parametrization of the dielectric function and sample reflectance calculated in the transfer-matrix formalism are plotted on top of the experimental data.
		}
	\label{figS2}
\end{figure}

To measure reflectance contrast we use a spectrally-broad tungsten-halogen lamp that is spatially filtered by an aperture and focused onto the sample by a glass-corrected 60x objective (NA=0.7) on an area of about 2\,$\mu$m diameter for whitelight (that is slightly larger than the 1.6\,$\mu$m spot size for the 532\,nm laser in the PL measurements).
We keep the total power on the order of a few 100's of nW or below to probe the optical response in the linear regime (for comparison we used 5\,$\mu$W power for PL).
The reflected light collected from the sample is guided through a grating spectrometer (300 gr/mm) and detected by a Peltier-cooled charge-coupled-device camera. 
The integration time for each frame reached from 0.3 to 1.8\,s to obtain values for maximum counts just below the saturation threshold, maximizing signal-to-noise.
The signal is further averaged over 10's up to 100's of individually acquired frames.
The data is then collected for each gate voltage, that is scanned in small steps, usually between 50 and 200 mV, between the measurements.
The voltage is further scanned in both directions acquiring several data sets at nominally equivalent external conditions to evaluate and exclude residual hysteresis effects (that are very small in our samples). 
The reflectance contrast is defined as 
\begin{equation}
\label{RC}
R_C=(R_{s}-R_{ref})/(R_{ref}-BG),
\end{equation}
where $R_{s}$ and $R_{ref}$ are sample and reference spectra measured with and without the WSe$_2$ monolayer, respectively, and $BG$ is the background signal acquired with the light-source switched off.
More specifically, following references were used depending on the sample structure and availability of respective regions: only Si, Si/hBN/hBN, and
Si/hBN/hBN/few-layer graphene.
Typical, as-measured $R_C$ data from a gated WSe$_2$ monolayer sample are presented in Fig.\,\ref{figS2} in the regime of excited state excitons as a color map in (a) and as individual, selected spectra in (b).
Here, multi-layer interference effects determine the overall lineshape of the resonances and the overall tilt of the spectra.
The excited state transition ($2s$) is detected at charge neutrality conditions with the $2s\pm$ resonances appearing at finite gate voltages.
Here, the sample is doped with free electrons and holes at positive and negative voltages, respectively. 
Due to the interferences we employ the transfer-matrix formalism (see Ref.\,\onlinecite{Byrnes2016,Raja2019,Brem2019}) to extract peak parameters, such as resonance energy, linewidth, and oscillator strength from the data.
The fits are adjusted to match the first derivative of the reflectance, as illustrated in Fig.\,\ref{figS2}\,(c). 
 
\subsection{Reproducibility across different devices}

\begin{figure}[ht]
	\centering
			\includegraphics[width=14 cm]{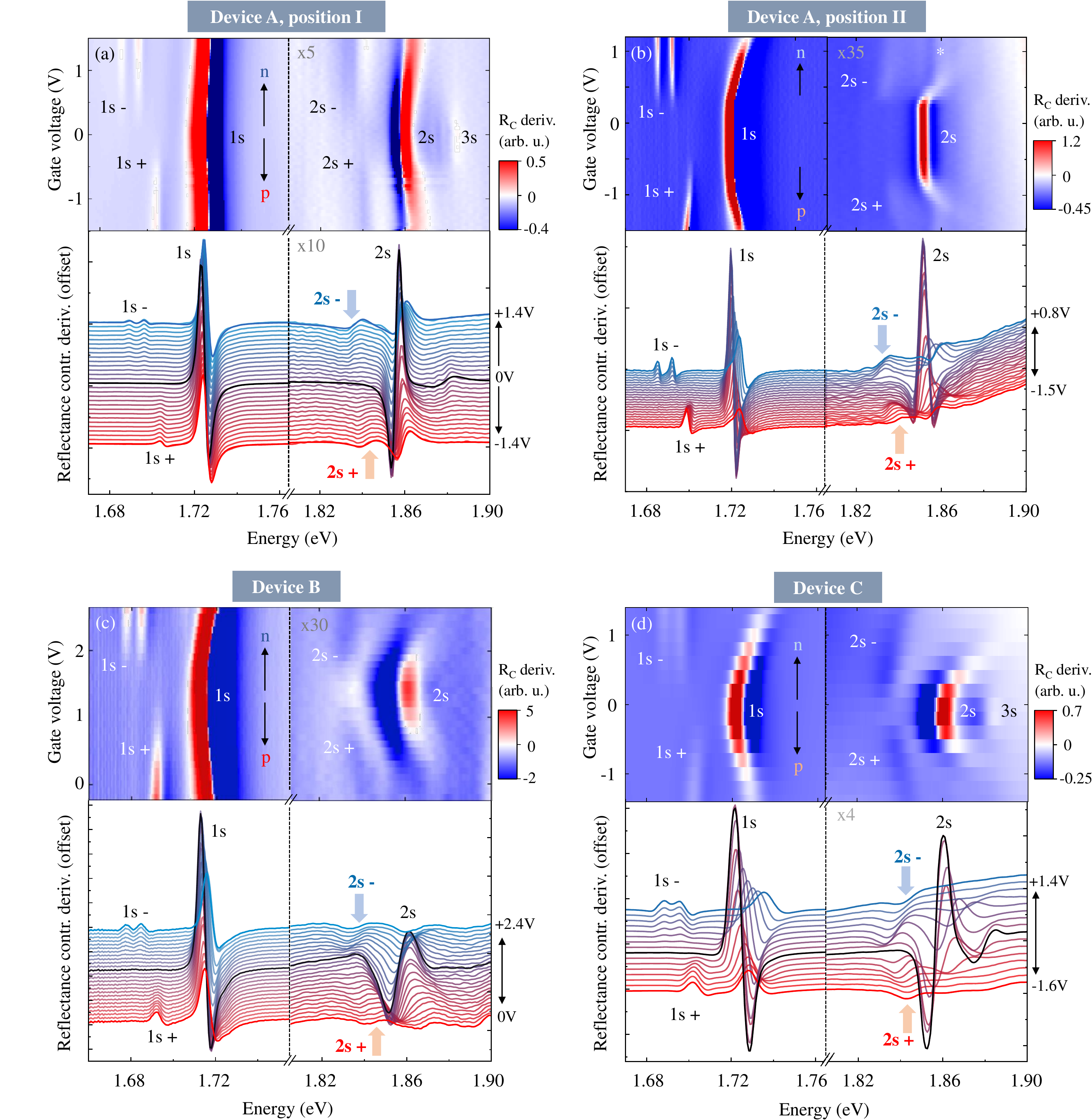}
		\caption{(a) Reflectance contrast first-derivative spectra of the sample position I on the device A are presented as a function of the gate voltage in the region of the ground ($1s$) and first excited ($2s$) exciton states.
		The lower panel shows selected spectra, vertically offset for clarity.
		(b)	Reflectance contrast derivatives for the sample position II on the same device.
		These measurements were used for the quantitative analysis of the majority of the data presented in Figs.~2 and 3 of the main text.
		(c) Same type of data for device B.
		(d) Same type of data for device C.
		For better visibility reflectance contrast derivatives in the excited state regime in the color maps are corrected by subtracting a baseline with a tilt. 
		}
	\label{figS3}
\end{figure}

Our measurements were performed on several devices and sample positions. 
Reflectance contrast derivatives from the device A on the sample position I are presented in the Fig.\,1 of the main text and reproduced in Fig.\,\ref{figS2}\,(a).
As further illustrated in the following Sec.\,\ref{density-voltage}, this particular part of the device allowed for a detailed study of a very low free carrier density regime, roughly between 10$^{10}$ and 10$^{11}$\,cm$^{-2}$.
For the quantitative analysis discussed in Figs.\,2 and 3 of the main text we used data collected on the same device at a sample position II that provided access to higher doping densities, up to several 10$^{12}$\,cm$^{-2}$.
Corresponding, as-measured contrast data are presented as color plot and selected spectra in Fig.\,\ref{figS3}\,(b) (these stem from the same data as presented in Fig.\,\ref{figS2}).
Additional data collected on a second (B) and third (C) devices are plotted in Figs.\,\ref{figS3}\,(c) and (d), respectively.

Altogether, we obtain very similar results in all studied cases regarding the electrical tunability of $1s$ and $2s$ resonances and, most importantly, the observation of $2s\pm$ resonances. 
Most notable differences between individual samples are related to linewidth broadening from residual inhomogeneities (e.g., compare devices A and B), since hBN may not always provide complete mitigation of environmental disorder\,\cite{Raja2019}, as well as the scaling of free carrier densities and all related effects with the gate voltage.
The latter is connected to both different thicknesses of the dielectric hBN layers as well as to the local nature and quality of the contacts. 
This point is discussed in more detail in the next Sec.\,\ref{density-voltage}. 

\section{Scaling of free career density with the gate voltage}
\label{density-voltage}

There are several ways to estimate free carrier densities in our measurements.
Very rough values can be given by considering the thickness of the gate dielectric and thus relate the applied voltage to the induced charge from the capacitance.
This approach, however, tends to be not too reliable in the low-density regime that is particularly susceptible to non-ohmic contacts and a generally non-linear dependence of carrier density on the gate voltage.
Alternatively, since the interactions of ground state excitons with free charges are well documented and partially understood at this stage, one can use a number of the spectroscopic observables for the same purpose.
These could be relative peak positions, linewidths, or oscillator strengths from the linear response that all scale linearly with the densities of free charge carriers (for densities well below the Mott-threshold for the $1s$ that is close to 10$^{13}$\,cm$^{-2}$).

This requires accurate quantitative measurements of these observables. Here, the relative energy shift $\Delta_{1s\pm}$ between $1s$ and $1s\pm$ is arguably one of the most convenient and robust metrics.
However, $\Delta_{1s\pm}$ scales with the Fermi-energy and thus density of the free charges with a specific pre-factor that is slightly different in the trion and Fermi-polaron models.
For trions it is close to the ratio between exciton and trion masses \,\cite{Esser2001} that is typically about 2/3 in TMDCs\,\cite{Kormanyos2014} and for Fermi-polarons it is the inverse of this ratio, thus in the range of 3/2\,\cite{Efimkin2017}.
Nevertheless, this approach has a clear advantage of being able to estimate the carrier density regardless of possible issues with local contacts, non-linear gate-voltage-density dependence or hysteresis.
It is particularly useful for samples and devices that offer a broad range of accessible doping density regions due to the difference both in the device structure (thickness of the hBN layers, e.g.) and local properties of the contacts.
We thus estimate the free carrier density according to $n_{e/h}=\Delta_{1s\pm} \times m_{e,h}/(\pi \hbar^2)$ by determining $\Delta_{1s\pm}$ from reflectance and setting it equal to Fermi-energy with a prefactor of unity for simplicity and consistency.
Due to the above issue with the scaling prefactor, absolute carrier densities estimated in this manner may deviate from the actual values by a constant factor from about 30\% to 50\%. 

\begin{figure}[ht]
	\centering
			\includegraphics[width=13 cm]{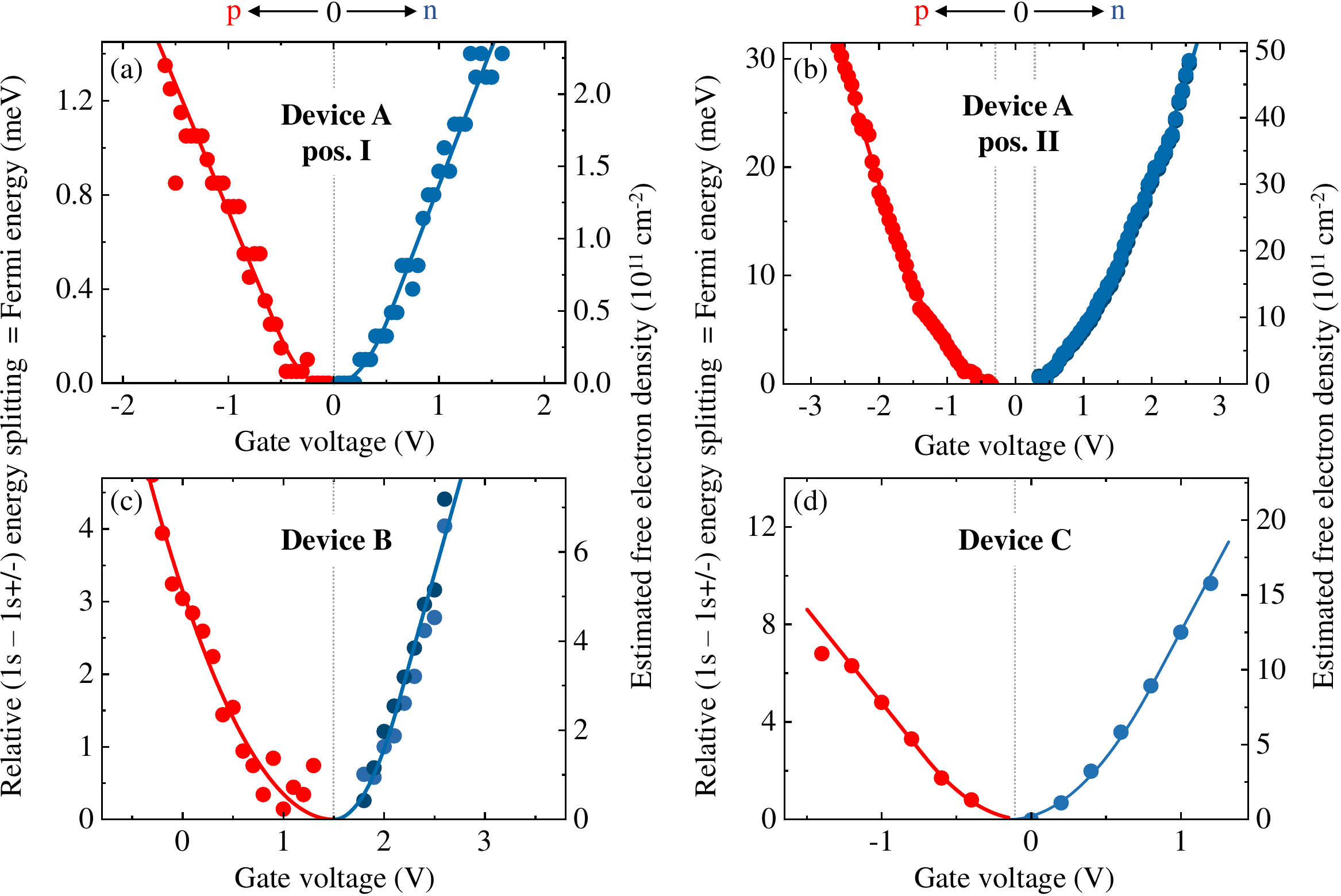}
		\caption{(a) - (d) Free-carrier doping densities extracted from the relative energy separation $\Delta_{1s\pm}$ between $1s$ and $1s\pm$ resonances across different devices and sample positions.
		According to both trion and Fermi-polaron models $\Delta_{1s\pm}$ is proportional up to a prefactor on the order of unity to the respective Fermi energy of the free-charge carriers.
		Right ordinate axis shows the resulting electron density.
		The hole density corresponds to $n_e$ multiplied by the hole-electron mass ratio $m_h/m_e$ that is about 0.9 in WSe$_2$ for the lowest bands\,\cite{Kormanyos2015}. 
		Calibration curves are shown by solid lines.
		}
	\label{figS4}
\end{figure}

Plots of $\Delta_{1s\pm} \approx E_F$ as function of applied gate-voltage and the estimated free electron densities are presented in Fig.\,\ref{figS4} for the studied sample positions and devices (corresponding to reflectance contrast data from Fig.\,\ref{figS3} in the previous section).
There are notable differences both in the measured dependence at very low gate voltages as well as in the absolute magnitude of the shifts.
For the analysis we use non-linear phenomenological fits to this data to obtain density-voltage calibration curves for each device that are shown in Fig.\,\ref{figS3} by solid lines.
Finally, we note that a higher gate voltages with approximately linear scaling this type of analysis is largely consistent with densities estimated from capacitance.

\section{Linewidths and oscillator strengths}
\subsection{Linewidths for extended free-carrier density range}

\begin{figure}[ht]
	\centering
			\includegraphics[width=11 cm]{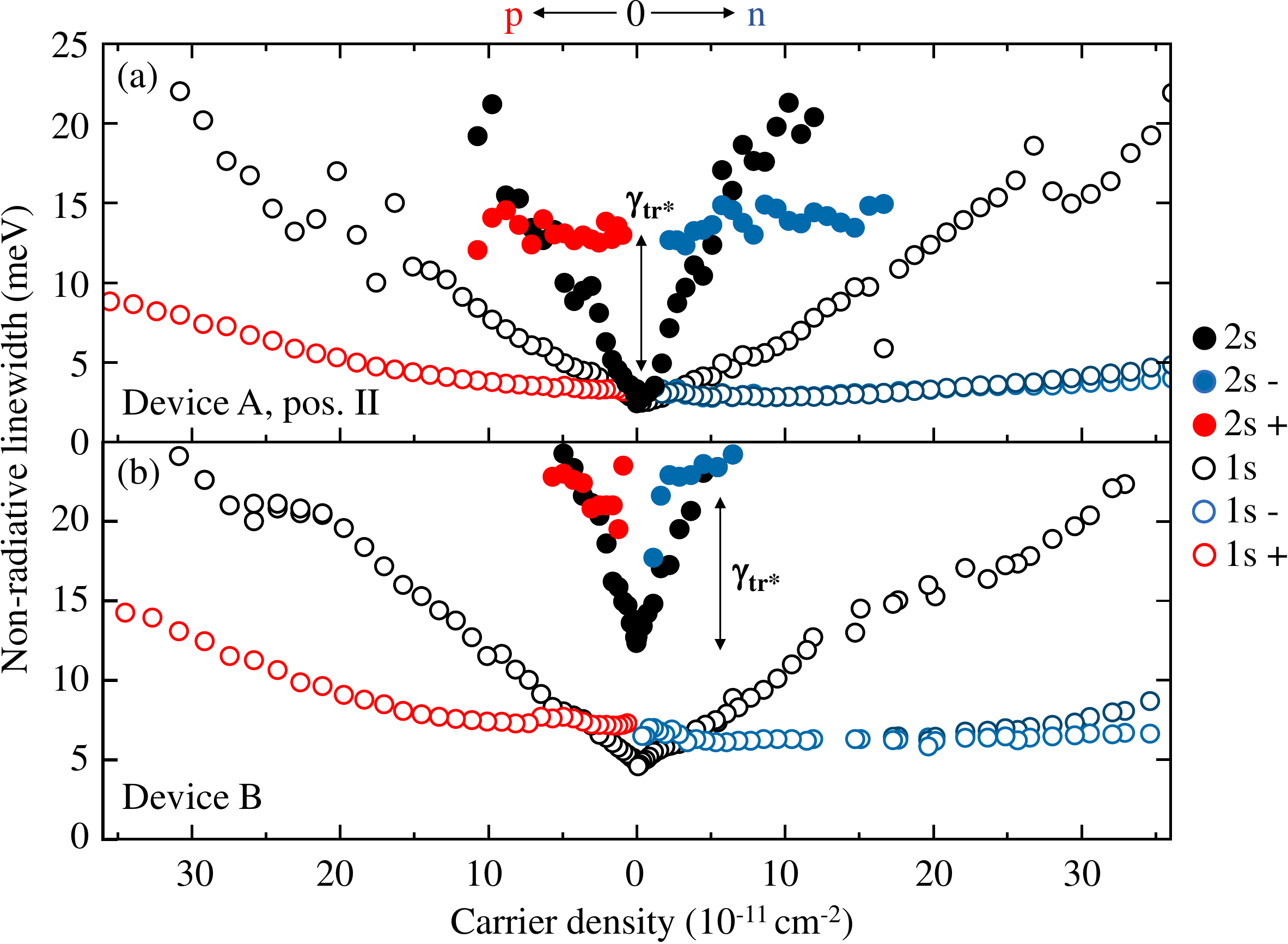}
		\caption{(a) Extracted non-radiative linewidths presented as full-width-at-half-maxima as function of free electron and hole densities for the ground ($1s$ and $1s\pm$) and first excited ($2s$ and $2s\pm$) states.
		The data corresponds to that presented in Fig.\,3 of the main text (device A, sample position II) and is presented across larger doping range.
		(b) Linewidths extracted from the data obtained on device B that was subject to residual disorder and resulting inhomogeneous broadening. 
		Characteristic broadening of the $2s\pm$ states from autoionisation is indicated by $\gamma_{tr,*}$.  
		}
	\label{figS5}
\end{figure}

Non-radiative linewidth from device A presented in the main text in Fig.\,3 are illustrated in Fig.\,\ref{figS5}\,(a) across a larger doping range, up to carrier densities of about $3.5\times10^{12}$\,cm$^{-2}$. 
Under these conditions the linewidths of $1s$ and $2s$ states continue to increase and also detect a weak density-induced broadening of the $1s\pm$ resonances that is absent at lower densities.
Interestingly, the broadening is more pronounced for $1s+$ rather than for $1s-$ states.
Corresponding linewidths extracted from device B are presented in Fig.\,\ref{figS5}\,(b).
This particular sample is subject to residual disorder and resulting inhomogeneous broadening that stems likely from the fluctuations in the dielectric environment\,\cite{Raja2019} due to a larger zero-density linewidths of $2s$ in contrast to $1s$.
Nevertheless, the two data sets demonstrate very similar results.
In particular, we consistently observe a characteristic, additional broadening $\gamma_{tr,*}$ of charged excited states in both cases.

\subsection{Oscillator strengths at low and high free-carrier densities}

\begin{figure}[h]
	\centering
			\includegraphics[width=10 cm]{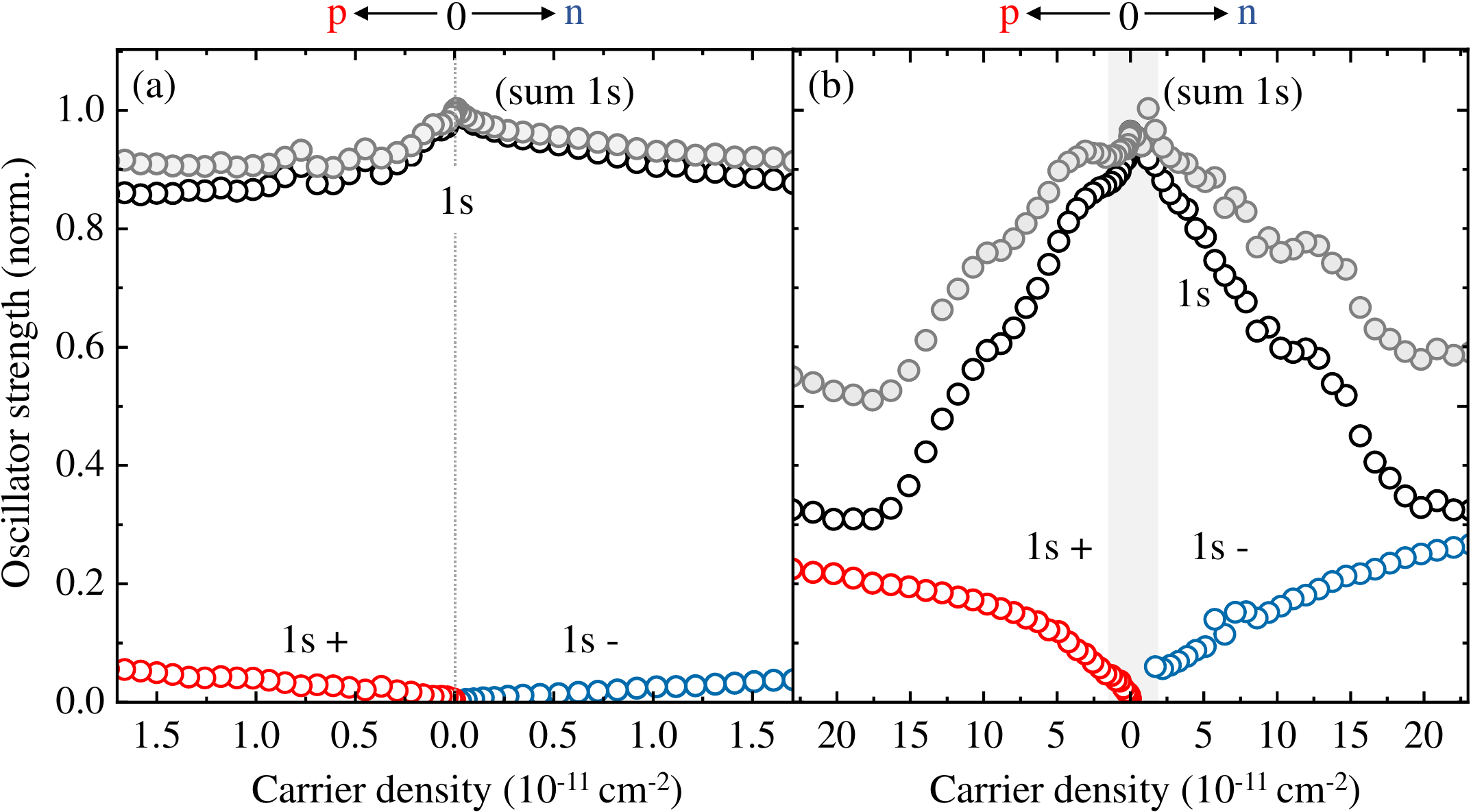}
		\caption{(a) Extracted oscillator strengths of the $1s$ and $1s\pm$ resonances in the very low doping range from a few $10^{10}$\,cm$^{-2}$ to $1.6\times10^{11}$\,cm$^{-2}$.
		(b) Corresponding data at elevated free carrier densities up to $2.3\times10^{12}$\,cm$^{-2}$.
		The combined oscillator strengths of $1s$ and $1s\pm$ is indicated by the gray circles in both plots.
		}
	\label{figS6}
\end{figure}

Additional data illustrating density-dependent oscillator strength of $1s$ and $1s\pm$ in very low and elevated doping range is presented in Fig.\,\ref{figS6} with (a) and (b) corresponding to sample positions I and II on the device A, respectively.
Total oscillator strengths from the sum of $1s$ and $1s\pm$ is plotted for comparison.
In both density regions we observe a gradual exchange of the oscillator strength between $1s$ and $1s\pm$. 
The two become roughly equal at about $2\times10^{12}$\,cm$^{-2}$.
This density threshold is almost an order of magnitude larger than that for the $2s$ state transferring half of the oscillator strength to $2s\pm$ (see Fig. 3\,(b) of the main text).
However, at very low doping conditions below $10^{11}$\,cm$^{-2}$ we detect a small additional decrease of the $1s$ oscillator strength that is not transferred to the respective charged ground states.
While the effect is rather small, on the order of 10\%, it is somewhat unexpected in a simple model.
Interestingly, due to the coupling between the exciton states via free charge carriers, parts of the oscillator strength can indeed be transferred from $1s$ to higher excited charged states, such as $2s$, $3s$, etc.
These processes are discussed theoretically in the following Sec.\,\ref{theory} and in particular in connection with the Eqs.\,\eqref{polaron:osc} and \eqref{polaron:osc:all}.
We note however, that both the assumptions of an analytical model and the difficulty of extracting absolute oscillator strengths of charged excited states at such very low densities currently preclude a more quantitative analysis.

\section{Photoluminescence and reflectance}

\begin{figure}[h]
	\centering
			\includegraphics[width=10.5 cm]{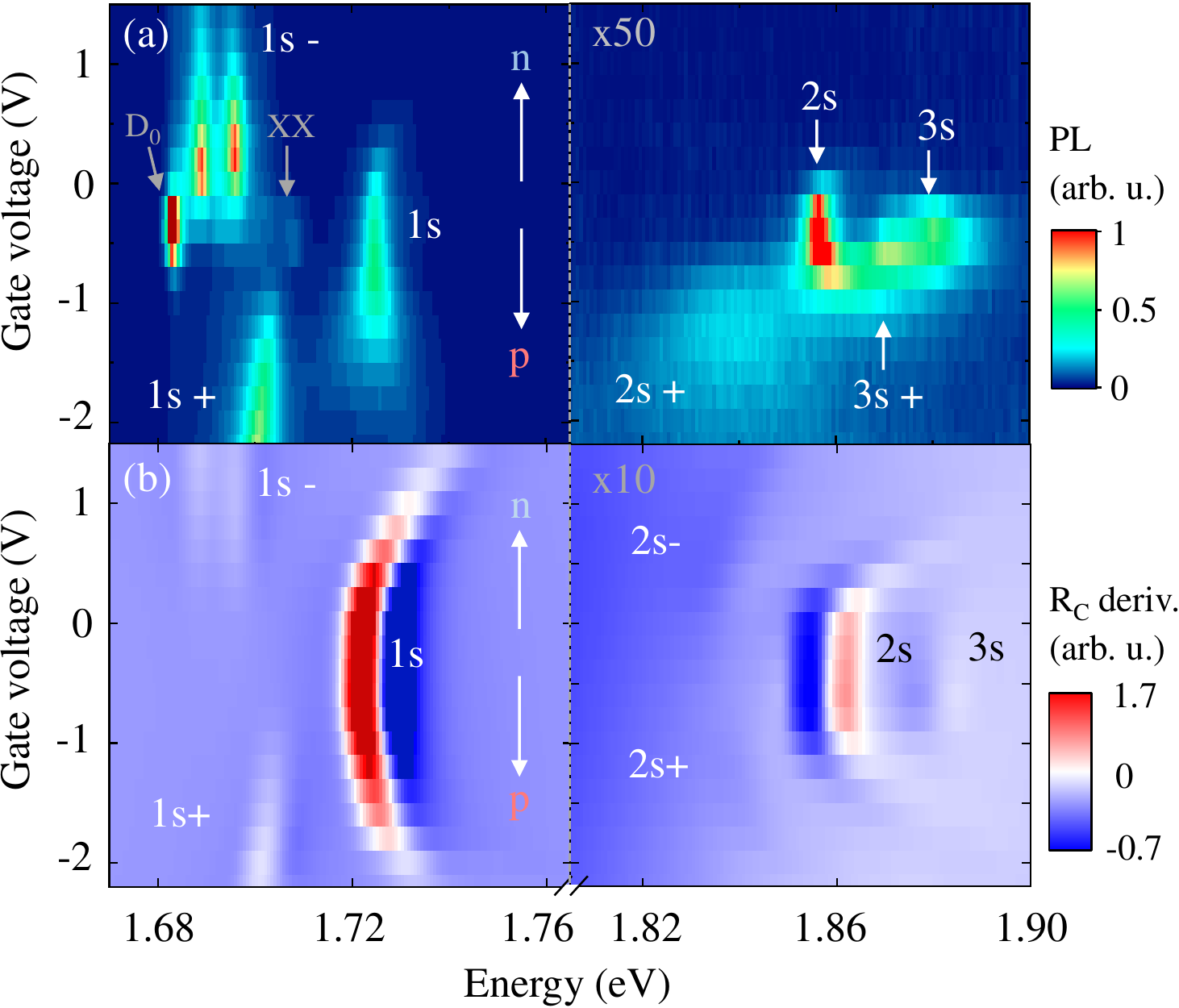}
		\caption{(a) Color plot of PL spectra as a function of gate voltage from continuous-wave excitation with a power density of 250\,W/cm$^2$ at the photon energy of 2.3\,eV, obtained on device C.
		(b) Corresponding reflectance contrast derivative spectra.
		The gate voltage of 1\,V corresponds to an estimated free carrier density on the order of $10^{12}$\,cm$^{-2}$.
		}
	\label{figS7}
\end{figure}

PL and reflectance contrast derivative spectra measured on device C are presented in Fig.\,\ref{figS7} as color plots in the range of $1s$ and $2s$ resonances.
As discussed in the main text we observe the emission from charged excited states corresponding to $1s\pm$ and $2s+$ resonances in reflectance. 
We note that while the reflectance probes the possibility to create these states directly by resonant excitation, the PL monitors actual populations that form after non-resonant injection.
In view of the ground state that is located more than 100\,meV below $2s$ and $2s\pm$, the emission from these transitions should be very far from the thermal equilibrium at 5\,K and occur during relaxation.

\begin{figure}[h]
	\centering
			\includegraphics[width=12 cm]{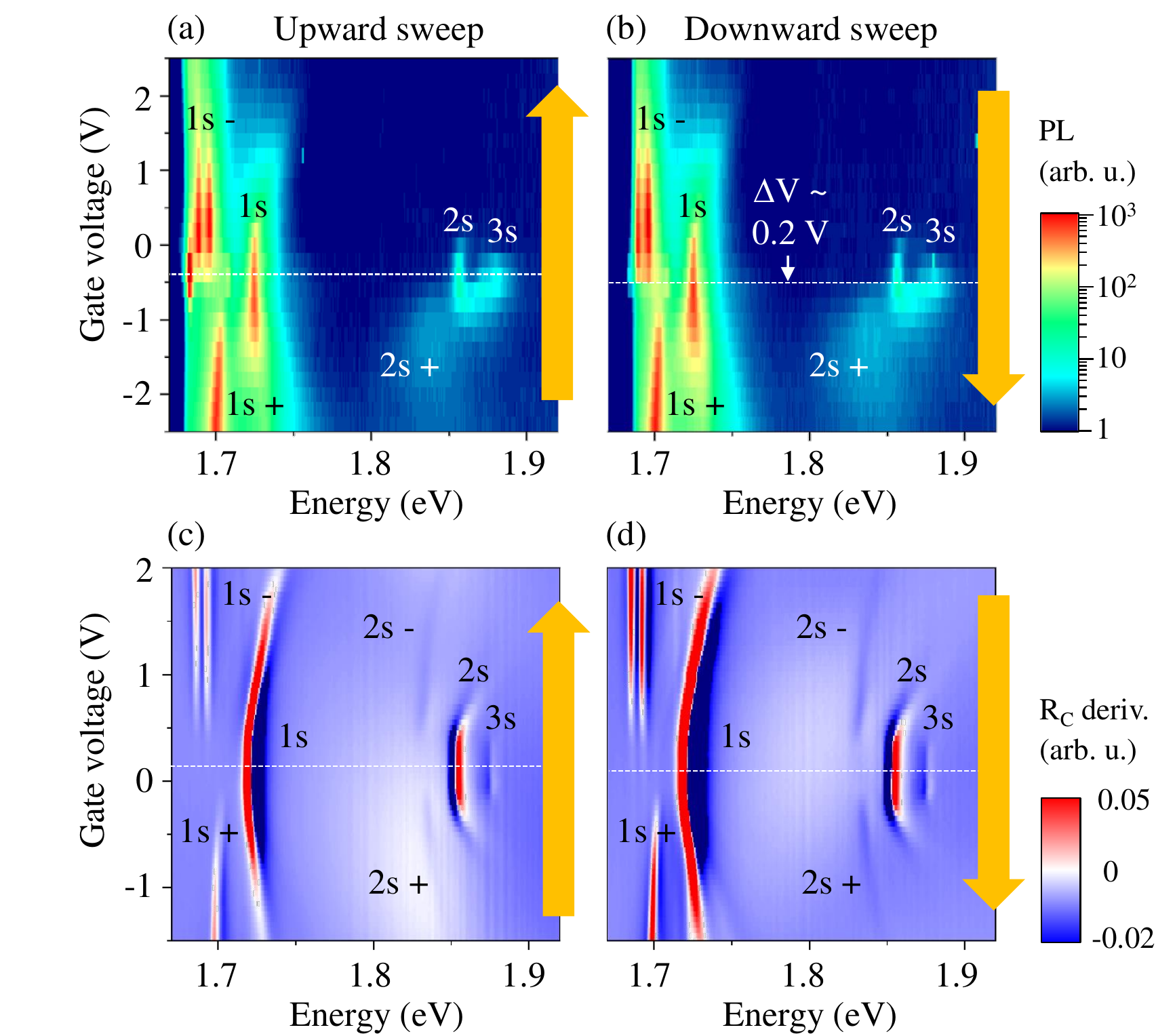}
		\caption{(a) Color plot of PL spectra as a function of gate voltage during the first sweep from -2.6 to +2.4\,V .
		(b) Corresponding data for the subsequent sweep from +2.6\,V to -3.2\,V.
		The PL counts are presented on a logarithmic scale for simultaneous comparison of the $1s$ and $2s$ intensity.
		Charge neutrality points are indicated by the dashed lines.
		(c) and (d) Reflectance contrast spectra for the two sweep directions obtained on a proximate sample position.
		}
	\label{figS8}
\end{figure}

For quantitative analysis we consider data acquired in the same sweep direction of the gate voltage and use the self-consistent doping-density estimation from spectroscopic observables, as outlined in Sec.\,\ref{density-voltage}.
One of the reasons for this approach is the presence of a small amount of hysteresis, typically below 10\% of the total doping range between upward and downward sweeps, as illustrated in Fig.\,\ref{figS8}.
Here we show two sets of PL spectra obtained during gate-voltage sweep from -2.6 to +2.4\,V (a) and then from +2.6\,V to -3.2\,V (b).
The hysteresis offset is estimated by comparing the behavior in the spectral range of $1s$ is on the order of 0.2\,V.
Similarly, the comparison of reflectance contrast spectra obtained on a proximate sample position exhibits only a small offset between upward and downward gate-voltage sweeps, as illustrated in Figs.\,\ref{figS8}\,(c) and (d).

\newpage
\section{Theoretical calculations of exciton-electron coupling}
\label{theory}
In this section we first outline general consequences of the exciton-electron interaction both for ground and excited states in a basic model of the short-range exciton-electron interaction. In particular, we provide additional details to the analysis and underlying derivations to the discussion presented in the main manuscript. Subsequently, we focus on a few more specific points concerning additional possible implications for peak positions and oscillator strengths as well as the influence of higher-excited exciton states. 
Here and in the following we note that the simplified analytical approach presented below involves approximations that may not fully represent the actual experimental scenarios, but is rather intended to transparently demonstrate the relevant general physics complementary to microscopic models. 

Below we derive the scattering matrix relevant for the electron-exciton interaction with allowance for the excited states and demonstrate that the excited states of charged excitons (trions) have intrinsic damping related to the autoionization effect. Further, using both Greens function and variational approach we derived dressed exction-electron states, Fermi-polarons, arising due to the coupling of excitons with the Fermi-sea. Finally, we briefly address the redistribution of the oscillator strengths.

\subsection{Derivation of the scattering amplitudes}

We apply the Fermi-polaron formalism~\cite{Suris2001,Suris2003,Sidler2016,Efimkin2017,Cotlet2019,Glazov2020a} for the excited trion features where an extra charge carrier is attached to the $2s$ exciton. We consider exciton-electron scattering in the short-range scattering model and introduce the matrix elements $V_{ij}$ for the exciton scattering from the state $j$ to $i$ (e.g., $V_{21}$ describes the scattering from $1s$ to $2s$ state). In our approach these matrix elements are the parameters of the theory and can be determined from the comparison with the experiment. The calculation of $V_{ij}$ is a separate problem beyond the scope of this paper, see Refs.~\cite{Fey2020,PhysRevB.65.153310,PhysRevB.67.045323} where the approaches to calculate $V_{ij}$ are developed. Let 
\begin{equation}
\label{greens:x}
G_j(\varepsilon, \bm k) = \frac{1}{\varepsilon - E_j - \frac{\hbar^2 k^2}{2m_x} + \mathrm i \Gamma_j},
\end{equation} 
be the exciton bare retarded Greens function, $E_j$ is its energy, $m_x$ is the exciton mass, $\Gamma_j$ is the damping of the state unrelated to the exciton-electron interaction. The equations for the scattering amplitudes $T_{ij}$ read (in the non-self-consistent approximation):
\begin{equation}
\label{scattering:T}
T_{ij}(\varepsilon,\bm k) = V_{ij} + \sum_l V_{il} \sum_{\bm p}(1-n_{\bm p}) G_l\left(\varepsilon - \frac{\hbar^2p^2}{2m_e}, \bm k - \bm p\right) T_{lj}(\varepsilon,\bm k),
\end{equation}
where $m_e$ is the electron mass, and $n_{\bm p}$ is the occupation of the electron state with momentum $\bm p$.

Equations~\eqref{greens:x} and \eqref{scattering:T} allow one to determine the trion -- exciton-electron correlated states in the non-self consistent approximation assuming that the electron-exciton interaction is sufficiently short-range (note that the electron-exciton interaction is discussed in detail in Ref.~\cite{Fey2020}). In general the subscripts $i,j,l$ run through all excitonic states, $1s$, $2s$, $2p$, etc., including the continuum states. We, however, restrict our model considering only several relevant states, particularly, $1s$ and $2s$.\footnote{Coupling between the $s-$ and $p-$ shell states (as well as with other states with the angular momentum component $l\ne 0$ is absent in the short-range interaction model where $V_{ij}$ are taken at the zero transferred wavevector due to the angular momentum conservation. We also note that due to the dependence of the observed resonance energies on the sign of the doping and their high oscillator strengths the states assigned as $2s\pm$ cannot be attributed to the $2p$ excitons which roughly fall in the same spectral range~\cite{2p:1,2p:2,PhysRevB.95.035311,Berghauser2016}.} Furthermore, to tackle the problem analytically we make the following simplifications. First of all, we neglect $n_{\bm p}$ in Eqs.~\eqref{scattering:T} assuming that the binding energies of excitons and trions exceed by far the electron Fermi energy~\cite{Glazov2020a}. The sum over $\bm p$ in Eq.~\eqref{scattering:T} can be readily calculated analytically with the result
\begin{equation}
\label{S:def}
S_l(\varepsilon,\bm k) = \sum_{\bm p} G_l\left(\varepsilon - \frac{\hbar^2p^2}{2m_e}, \bm k - \bm p\right) = \sum_{\bm p} \frac{1}{\varepsilon - E_l - \frac{\hbar^2p^2}{2\mu_{eX}}  -\frac{\hbar \bm k\cdot \bm p}{m_x} - \frac{\hbar^2 k^2}{2m_x} + \mathrm i \Gamma_l},
\end{equation} 
where $\mu_{eX}^{-1}= m_e^{-1} + m_x^{-1}$ is the exciton-electron reduced mass. We further put $\bm k=0$ (since we will need optically active states within the light cone), thus\footnote{We set the normalization area to unity and measure the matrix elements in the units of inverse density of states.}
\begin{equation}
\label{S:1}
S_l(\varepsilon,0) = \sum_{\bm p} \frac{1}{\varepsilon - E_l - \frac{\hbar^2p^2}{2\mu_{eX}}  + \mathrm i \Gamma_l} = \mathcal D \int_0^\infty \frac{d E_p}{\varepsilon - E_l - E_p + \mathrm i \Gamma_l} \approx \mathcal D\ln\left(\frac{E_l  - \varepsilon - \mathrm i \Gamma_{l}}{\bar E} \right),
\end{equation}
with $\mathcal D = \mu_{eX}/(2\pi \hbar^2)$, $\bar E$ being the cut-off energy. In what follows we are interested in the energy range where $\varepsilon \approx E_{2s}$ (spectral vicinity of the excited, $2s$, exciton and corresponding trion), thus we set
\begin{equation}
\label{S:2}
S_1(\varepsilon,0) = -\mathrm i \pi \mathcal D, \quad S_2(\varepsilon,0) = \mathcal D\ln\left(\frac{E_{2s}  - \varepsilon - \mathrm i \Gamma}{\bar E}\right).
\end{equation}
It follows from Eq.~\eqref{scattering:T} that
\begin{equation}
\label{scattering:T:1}
T_{ij} = V_{ij} + \sum_l V_{il} S_l T_{lj},
\end{equation}
i.e., the scattering amplitudes satisfy the set of algebraic equations which can be readily solved analytically in the case of two relevant states and, in general, numerically.

Following the standard approach, where the trion state can be considered as an electron loosely bound to a rigid exciton, we assume that the bare matrix elements are small
\[
\mathcal D|V_{ij}| \ll 1.
\]
Let us start with the analysis of the amplitudes $T_{i2}$ relevant for the $2s$ exciton. Under the assumptions above, the scattering amplitudes $T_{22}$, $T_{12}$ satisfy the equations
\begin{equation}
\label{T22}
T_{22} = (1+S_2 T_{22}) \left[V_{22} +\frac{|V_{12}|^2 S_1}{1-V_{11} S_1} \right], \quad T_{12} = \frac{V_{12}(1+S_2 T_{22}) }{1-V_{11} S_1}
\end{equation}
As a result, one can introduce the effective excited exciton-electron interaction parameter
\begin{equation}
\label{V22:t}
\tilde V_{22} = V_{22} +\frac{|V_{12}|^2 S_1}{1-V_{11} S_1} \approx V_{22} - \mathrm i |V_{12}|^2 \pi \mathcal D.
\end{equation}
and express
\begin{equation}
\label{T22:1}
T_{22} = \frac{1}{\tilde V_{22}^{-1} - S_2} = \frac{1}{\mathcal D} \frac{1}{\ln\left[\frac{\bar E}{E_{2s} - \varepsilon} \exp{\left(\frac{1}{\mathcal D V_{22} - \mathrm i |V_{12}|^2 \pi \mathcal D^2} \right)} \right]}.
\end{equation}
The difference with the standard Fermi-polaron model is the presence of imaginary part in $\tilde V_{22}$ due to the process 
\[
(2s~\mbox{exciton}+\mbox{electron})_{bound} \to 1s~\mbox{exciton} + \mbox{electron}'_{free},
\]
see the main text for illustration.
Equation~\eqref{T22} has a pole corresponding to the ``excited trion resonance'' at 
\begin{equation}
\label{2s:trion}
\varepsilon = E_{2s} - \bar E \exp{\left(\frac{V_{22}+\mathrm i |V_{12}|^2\pi \mathcal D}{\mathcal D(|V_{22}|^2 + |V_{12}|^4\pi^2\mathcal D^2)} \right)}  = E_{2s} - E_{tr,2s}
\left(1+ \mathrm i \tan{\phi}\right),
\end{equation}
where
\begin{equation}
\label{binding}
E_{tr,2s} = \bar E \exp{\left(\frac{V_{22}}{\mathcal D(|V_{22}|^2 + |V_{12}|^4\pi^2\mathcal D^2)} \right)} \cos{\phi} \approx \bar E \exp{\left(\frac{1}{\mathcal D V_{22}} \right)}\cos{\phi}, 
\end{equation}
is the excited trion binding energy ($V_{22}<0$), and
\[
\phi = \frac{\pi |V_{12}|^2}{|V_{22}|^2 + \pi^2\mathcal D^2 |V_{12}|^4} \approx \pi  \frac{ |V_{12}|^2}{|V_{22}|^2}.
\]
The ``intrinsic'' trion damping reads
\begin{equation}
\label{damping}
\gamma_{tr} \approx \bar E \exp{\left(\frac{1}{\mathcal D V_{22}} \right)}\sin{\phi} = E_{tr,2s} \tan{\phi}.
\end{equation}
Equations~\eqref{binding} and \eqref{damping} are in agreement with Eqs. (5) and (7) of the main text.

Note that $\phi$ should not exceed $\pi/2$ ($|V_{12}| \lesssim |V_{22}|$), otherwise the model fails since trion binding energy becomes negative.\footnote{It opens the way to control the excited state trions by varying $V_{12}$, e.g., by fine tuning of the Coulomb interaction in van der Waals heterostructures. The fact that $\phi>\pi/2$ might possibly be the reason why the excited trions are not observed for conventional quantum well structures.} In the vicinity of the pole 
\begin{equation}
\label{T22:pole}
T_{22} \approx \frac{1}{\mathcal D} \frac{1}{\ln{\left(\frac{E_{tr,2s}(1+\mathrm i \tan{\phi})}{E_{2s}-\varepsilon}\right)}} \approx \frac{E_{tr,2s}}{\mathcal D} \frac{1}{ \varepsilon + E_{tr,2s} + \mathrm i \gamma_{tr} - E_{2s}}, \quad T_{12} \approx \frac{V_{12}}{V_{22}} T_{22}.
\end{equation}
Analogously, 
\begin{equation}
\label{T11}
T_{11} = (1+S_1 T_{11}) \left[V_{11} +\frac{|V_{12}|^2 S_2}{1-V_{22} S_2} \right], \quad T_{21} = \frac{V_{21}(1+S_1 T_{11}) }{1-V_{22} S_2}.
\end{equation}
We are interested in the features in $T_{22}$ in the vicinity of $\varepsilon \approx E_{2s} - E_{tr,2s}$. Accordingly, we recast
\begin{equation}
\label{T11:1}
T_{11} = \frac{1}{\left(V_{11} +\frac{|V_{12}|^2 S_2}{1-V_{22} S_2}\right)^{-1} - S_1} \approx \frac{|V_{12}|^2}{V_{22}^2} T_{22}, \quad T_{21} \approx \frac{V_{21}}{V_{22}} T_{22}.
\end{equation}
The scattering amplitudes determine the exciton-electron bound or correlated states in the limit of zero doping. Our next step is to account for the finite doping and determine the Fermi-polaron resonances, i.e., the dressed exciton-electron states.

\subsection{Fermi-polarons Greens functions}

We take the self-energies in the simplest form~\cite{Glazov2020a}
\begin{equation}
\label{self-gen}
\Sigma_{ij} = N_e T_{ij}.
\end{equation} 
Thus, the matrix Greens function $\mathcal G_{ij}$ of the excitons dressed by the Fermi-sea satisfies the following equations
\begin{equation}
\label{GG:set}
\mathcal G_{ij} = G_i \delta_{ij} + N_e G_i T_{il} \mathcal G_{lj},
\end{equation}
therefore
\begin{subequations}
\label{GG:sol}
\begin{align}
\mathcal G_{11} = \frac{1}{G_1^{-1}-N_{e} T_{11} - N_e^2\frac{T_{12}T_{21} }{G_2^{-1}-N_e T_{22}}},\label{GG11}\\
\mathcal G_{22} = \frac{1}{G_2^{-1}-N_{e} T_{22} - N_e^2\frac{T_{12}T_{21} }{G_1^{-1}-N_e T_{11}}}.
\label{GG22}
\end{align}
\end{subequations}
These expressions can be recast in somewhat different form
\begin{subequations}
\label{GG:sol:1}
\begin{equation}
\label{GG11:1}
\mathcal G_{11} = \frac{G_2^{-1}-N_e T_{22}}{
(G_1^{-1}-N_{e} T_{11})(G_2^{-1}-N_e T_{22}) - N_e^2 T_{12}T_{21}
},
\end{equation}
\begin{equation}
\label{GG22:1}
\mathcal G_{22} = \frac{G_1^{-1}-N_e T_{11}}{
(G_1^{-1}-N_{e} T_{11})(G_2^{-1}-N_{e} T_{22}) - N_e^2T_{12}T_{21} 
}.
\end{equation}
\end{subequations}
The Fermi-polarons appear as poles of the Greens functions (naturally, the positions of poles in $\mathcal G_{11}$ and $\mathcal G_{22}$ coincide):
\begin{empheq}{equation}
\label{polarons}
(\varepsilon - E_{1s} +\mathrm i \Gamma_{1s}  - N_e T_{11})(\varepsilon - E_{2s} +\mathrm i \Gamma_{1s}  - N_e T_{22}) - N_e^2 T_{12}T_{21}=0.
\end{empheq}
We recall that $T_{ij}$ have poles at the energies of the trion resonances, and it follows from Eq.~\eqref{polarons} that the coupling between the $1s$ and $2s$ excitons and polarons arises $\propto N_e^2 T_{12} T_{21}$.
We can recast Eq.~\eqref{polarons} in somewhat different form by virtue of approximate explicit pole-like forms of $T_{ij}$, Eqs.~\eqref{T22:pole} and \eqref{T11:1}, useful for qualitative analysis and comparison with other approaches:
\begin{multline}
\label{polarons:1}
\left(\varepsilon - E_{1s} +\mathrm i \Gamma_{1s}  - N_e  \frac{|V_{12}|^2}{V_{22}^2} \frac{E_{tr,2s}/\mathcal D}{\varepsilon - E_{2s} + E_{tr,2s} + \mathrm i \gamma_{tr} }\right)\left(\varepsilon - E_{2s} +\mathrm i \Gamma_{1s}  - N_e \frac{E_{tr,2s}/\mathcal D}{\varepsilon - E_{2s} + E_{tr,2s} + \mathrm i \gamma_{tr} }\right) \\
- N_e^2 \frac{|V_{12}|^2}{V_{22}^2} \left(\frac{E_{tr,2s}/\mathcal D}{\varepsilon - E_{2s} + E_{tr,2s} + \mathrm i \gamma_{tr} }\right)^2 =0.
\end{multline}
We analyze the solutions of Eq.~\eqref{polarons:1} in Sec.~\ref{subsec:FB:analysis}. Below we derive Eqs.~\eqref{polarons:1} by the variational approach for the Fermi-polaron.

For a variational calculation we employ the Chevy ansatz~\cite{Sidler2016,PhysRevA.74.063628} (as before, we set the exciton momentum to $0$, $i,j=1,2$ enumerate excitonic states; we also disregard the damping of the states for simplicity):
\begin{equation}
\label{chevy}
\Psi = \sum_{i} \varphi_i X^\dag_{i,0}|0\rangle + \sum_{j,\bm p, \bm q} F_{j,\bm p,\bm q} X^\dag_{j,\bm q-\bm p} e^\dag_{\bm p} e_{\bm q}|0\rangle.
\end{equation}
Here $X^\dag$, $X$ are the exciton creation and annihilation operators, $e^\dag$, $e$ are the electron creation and annihilation operators, $\varphi_i$ is the amplitude of the excitonic state in the polaron, $F_{j,\bm p,\bm q}$ is the amplitude of the correlated state. The state $|0\rangle$ corresponds to the Fermi see, accordingly, in Eq.~\eqref{chevy} $q<k_F$. In what follows in all summations for observables the condition $q<k_F$ is implied. The energy is evaluated as follows
\[
\mathcal E = \langle \Psi|\mathcal H|\Psi\rangle =  K+  P,
\]
where 
\begin{equation}
\label{Hamiltonian}
\mathcal H = \sum_{\bm k} \left(E_i+ \frac{\hbar^2k^2}{2m_x}\right) X^\dag_{i,\bm k}X_{i,\bm k} + \sum_{\bm p} \frac{\hbar^2 p^2}{2m_e} e^\dag_{\bm p} e_{\bm p} + \sum_{ij,\bm k, \bm p,\bm q} V_{ij} X_{i,\bm k+\bm p}^\dag X_{j,\bm k} e^\dag_{\bm q-\bm p} e_{\bm q}. 
\end{equation}
The energy should be minimized with respect to $\varphi_i$ and $F_{j,\bm p,\bm q}^i$.
Kinetic energy reads
\begin{equation}
\label{chevy:K}
K = \sum_i E_i |\varphi_i |^2 + \sum_{j,\bm p, \bm q} |F_{j,\bm p,\bm q}|^2 \left(E_j + \frac{\hbar^2 (\bm q - \bm p)^2}{2m_x} + \frac{\hbar^2 p^2}{2m_e} - \frac{\hbar^2 q^2}{2m_e} \right).
\end{equation}
Potential energy takes the form
\begin{equation}
\label{chevy:P}
P = \sum_{ij,\bm q} V_{ij} \varphi_i^*\varphi_j + \sum_{ij,\bm k, \bm q, \bm p} V_{ij} F_{i,\bm k,\bm q}^* F_{j,\bm p,\bm q} + \sum_{ij,\bm k, \bm q, \bm q'} V_{ij} F_{i,\bm k,\bm q}^* F_{j,\bm k,\bm q'}
+ \sum_{ij,\bm k, \bm q} V_{ij} \varphi_i^* F_{j,\bm k,\bm q} + \sum_{ij,\bm k, \bm q}  V_{ij} F_{i,\bm k,\bm q}^* \varphi_j,
\end{equation}
the first term here arises due to $\bm q=0, \bm k=0$ in term in Eq.~\eqref{Hamiltonian}.

Variational derivatives $\delta \mathcal E/\delta\varphi_i^*$ and $\delta \mathcal E/\delta F^*_{i,\bm k,\bm q}$ can be readily calculated and allow us to obtain the equations for $\varphi_i$, $F_{i,\bm k,\bm q}$:
\begin{subequations}
\begin{align}
E_i \varphi_i + \sum_{j,\bm q} V_{ij} \varphi_j + \sum_{j,\bm k,\bm q} V_{ij} F_{j,\bm k, \bm q} = \varepsilon \varphi_i,\label{eq:phi}\\
\left(E_j+ \frac{\hbar^2 (\bm q - \bm k)^2}{2m_x} + \frac{\hbar^2 k^2}{2m_e} - \frac{\hbar^2 q^2}{2m_e} \right) F_{i,\bm k, \bm q} + \sum_{j} V_{ij}\varphi_j + \sum_{j,\bm p} V_{ij} F_{j,\bm p, \bm q} =\varepsilon F_{i,\bm k,\bm q}.\label{eq:F}
\end{align}
\end{subequations}
According to Refs.~\cite{Sidler2016,PhysRevA.74.063628} one can safely disregard the contribution $\propto \sum_{\bm q'} F_{j,\bm k,\bm q'}$ (omitted in the last equation) since this contribution is convergent, in contrast to the contributions with $\sum_{\bm k} F_{j,\bm k,\bm q}$ which formally diverges at high wavevectors.
We introduce auxiliary functions
\begin{equation}
\label{aux}
\chi_i (\bm q) = \varphi_i + \sum_{\bm k} F_{i,\bm k,\bm q},
\end{equation}
and express $F_{i,\bm k,\bm q}$ from Eq.~\eqref{eq:F} as
\begin{equation}
\label{Fi:1}
F_{i,\bm k,\bm q} = \sum_{j} \frac{V_{ij} \chi_j(\bm q)}{\varepsilon - \left(E_j+ \frac{\hbar^2 (\bm q - \bm k)^2}{2m_x} + \frac{\hbar^2 k^2}{2m_e} - \frac{\hbar^2 q^2}{2m_e} \right)}.
\end{equation}
Summing Eq.~\eqref{Fi:1} over $\bm k$ and making use of the definition~\eqref{aux} we obtain
\begin{equation}
\label{chi}
\chi_i(\bm q) = \varphi_i + \sum_{j,\bm k} \frac{V_{ij} \chi_j(\bm q)}{\varepsilon - \left(E_j+ \frac{\hbar^2 (\bm q - \bm k)^2}{2m_x} + \frac{\hbar^2 k^2}{2m_e} - \frac{\hbar^2 q^2}{2m_e} \right)} = \varphi_i + \sum_j V_{ij} S_j \chi_j(\bm q).
\end{equation}
Here the functions $S_i$ are introduced in Eq.~\eqref{S:1} and we applied the same assumptions as above regarding the kinetic energy and recoil (i.e., we set $\bm q=0$, in what follows the summation over $\bm q$ and $q<k_F$ is replaced by the multiplication by $N_e$). In order to establish the relation with above we note that ($\chi$ and $\varphi$ denote the column vectors $[\chi_1,\chi_2]^T$, $[\varphi_1,\varphi_2]^T$, respectively, $\hat V$, $\hat T$ and $\hat S$ denote corresponding matrices)
\begin{equation}
\label{chi:1}
\chi = (1-\hat V \hat S)^{-1} \varphi, \quad \hat T= (1-\hat V \hat S) \hat V \quad \Rightarrow \quad \chi = \hat V^{-1} \hat T  \varphi.
\end{equation}

Now we are able to find the energy in the standard way from Eq.~\eqref{eq:phi}. Taking into account that, under the assumptions above, the summation over $\bm q$ ($q<k_F$) is equivalent to multiplication by the electron density $N_e$ we obtain the following matrix equation for $\varphi$:
\begin{equation}
\label{eq:phi:1}
\begin{pmatrix}
E_1 & 0\\
0 & E_2
\end{pmatrix}\varphi + N_e\hat V \chi = \begin{pmatrix}
E_1 & 0\\
0 & E_2
\end{pmatrix}\varphi + N_e\hat T \varphi = \begin{pmatrix}
\varepsilon & 0\\
0 & \varepsilon
\end{pmatrix}\varphi,
\end{equation}
or
\begin{equation}
\label{eq:phi:2}
\begin{pmatrix}
E_1 + N_e T_{11}& N_e T_{12}\\
N_e T_{21} & E_2 + N_2 T_{22}
\end{pmatrix}\varphi  = \begin{pmatrix}
\varepsilon & 0\\
0 & \varepsilon
\end{pmatrix}\varphi.
\end{equation}
The equation for the eigenvalues is exactly the same as Eq.~\eqref{polarons} at $\Gamma \to 0$.

\subsection{Fermi polarons. Analysis of the results}\label{subsec:FB:analysis}

Let us analyze the solutions of Eq.~\eqref{polarons:1} in detail. Since we are interested in $\varepsilon$ close to $E_{2s} - E_{tr,2s}$, we can rearrange this equation by multiplying it by $(\varepsilon - E_{2s} + E_{tr,2s} + \mathrm i \gamma_{tr})^2$ and dividing by $(\varepsilon - E_{1s} +\mathrm i \Gamma_{1s})(\varepsilon - E_{2s} +\mathrm i \Gamma_{1s}) \approx -(E_{2s} - E_{1s} - E_{tr,2s})E_{tr,2s}$. As a result, we obtain the quadratic equation with the roots
\begin{subequations}
\label{FP:roots}
\begin{align}
&\varepsilon = E_{2s} - E_{tr,2s},\label{spur:root}\\
&\varepsilon = E_{2s} - E_{tr,2s} - N_e\frac{E_{tr,2s}}{\mathcal D}\left(\frac{1}{E_{tr,2s}} - \frac{|V_{12}|^2}{V_{22}^2}\frac{1}{E_{2s}-E_{1s} -E_{tr,2s}}\right).\label{FP:root}
\end{align}
\end{subequations}
In fact, the ``bare'' trion energy, Eq.~\eqref{spur:root}, is a spurious solution, thus pole is actually absent in the Greens functions, which, in the vicinity of $\varepsilon \approx E_{2s} - E_{tr,2s}$ take the form:
\begin{subequations}
\label{Greens:FP}
\begin{align}
\mathcal G_{22} \approx \frac{N_e/(E_{tr,2s}\mathcal D)}{\varepsilon - E_{2s} + E_{tr,2s} + \mathrm i \gamma_{tr} - N_e\frac{E_{tr,2s}}{\mathcal D}\left(\frac{1}{E_{tr,2s}} - \frac{|V_{12}|^2}{V_{22}^2} \frac{1}{E_{2s}-E_{1s} -E_{tr,2s}}\right)},\label{G22:FP}\\
\mathcal G_{11} \approx \frac{|V_{12}|^2}{V_{22}^2} \left(\frac{E_{tr,2s}}{E_{1s} - E_{2s} + E_{tr,2s}} \right)^2\mathcal G_{22}.\label{G11:FP}
\end{align}
\end{subequations}
Thus, in the optical susceptibility of the TMDC ML in the vicinity of the excited trion energy there is a pole at [see Eq.~\eqref{FP:root}]
\begin{subequations}
\label{polarons:1+2}
\begin{empheq}{equation}
\label{polaron:res}
\hbar \omega = E_{2s} - E_{tr,2s} - N_e\frac{E_{tr,2s}}{\mathcal D}\left(\frac{1}{E_{tr,2s}} - \frac{|V_{12}|^2}{V_{22}^2}\frac{1}{E_{2s}-E_{1s} -E_{tr,2s}}\right).
\end{empheq}
This resonance acquires oscillator strength from both the ground, $1s$, and excited, $2s$, excitons
\begin{empheq}{equation}
\label{polaron:osc}
f = \frac{N_e}{\mathcal D}\left[f_{1s} \frac{|V_{12}|^2}{V_{22}^2} \frac{E_{tr,2s}}{(E_{1s} - E_{2s} + E_{tr,2s})^2} + f_{2s}\frac{1}{E_{tr,2s}}\right].
\end{empheq}
\end{subequations}

Equations~\eqref{polarons:1+2} can be extended to account for the coupling with the whole series of the excited states as follows:\footnote{Equations~\eqref{polarons:1+2+all} can be also derived using the method of Ref.~\cite{imamoglu2020excitonpolarons}}
\begin{subequations}
\label{polarons:1+2+all}
\begin{empheq}{equation}
\label{polaron:res:all}
\hbar \omega = E_{2s} - E_{tr,2s} + N_e\frac{E_{tr,2s}}{\mathcal D}\frac{|V_{n2}|^2}{V_{22}^2} \sum_n\frac{1}{E_{2s} - E_{tr,2s} - E_{ns}},
\end{empheq}
and
\begin{empheq}{equation}
\label{polaron:osc:all}
f = N_e\frac{E_{tr,2s}}{\mathcal D}\sum_n f_{ns} \frac{|V_{n2}|^2}{V_{22}^2} \frac{1}{(E_{ns} - E_{2s} + E_{tr,2s})^2} .
\end{empheq}
\end{subequations}

For crude estimations of the effect of the higher excited states, the $n$-dependence of the denominator can be disregarded assuming
\[
E_{ns} - E_{2s} + E_{tr,2s} \approx E_{tr,2s},
\]
because the corrections due to $E_{ns}$ and $E_{2s}$ are $\propto 1/n^2$.\footnote{Note that for excited states the hydrogenic description is appropriate due to effective averaging of the dielectric functions of the environment~\cite{PhysRevMaterials.2.011001}.} There are two relevant mechanisms of the electron-exciton interaction contributing to $V_{ij}$~\cite{Fey2020,PhysRevB.65.153310,PhysRevB.67.045323}: the exchange (and, possibly, direct) Coulomb interaction and the polarization interaction. The latter should dominate for high-$n$ states due to their enhanced polarizability. For polarization  interaction (conserving the principal quantum number $n$) $V(r) \propto \alpha_n/r^4$, where $\alpha_n$ is the polarizability of the state which scales as $\alpha_n \propto n^7$~\cite{RevModPhys.82.2313,PhysRevB.96.125142}.  Thus, one may crudely estimate $|V_{n2}|^2 \propto n^7$ which results in divergence in Eq.~\eqref{polaron:res:all} and also in Eq.~\eqref{polaron:osc:all} ($f_{ns} \propto 1/n^3$). Similar crude estimates for the exchange interaction provide $|V_{n2}|^2 \sim E_{B,n} a_{B,n}^2 E_{B,2} a_{B,2}^2 \propto n^2$, that is why the power-law divergence of the polaron shift and logarithmic divergence of the oscillator strength are also expected. It means that the contribution of the excited exciton states to the Fermi-polaron is expected to be very important, but the perturbative treatment is not sufficient.

That is why we abstain from the analysis of the redistribution of the oscillator strength between the excited exciton and corresponding Fermi-polaron resonance. Additionally, the prerequisite of the model, the inequality 
\begin{equation}
\label{pre:ineq}
\mathcal D |V_{ij}| \ll 1
\end{equation}
is not satisfied for the excited states where, as it follows from the experiment, $\mathcal D |V_{22}| \approx 1$ and $E_{tr,2s} \sim E_{2s}$. This renders perturbative description of the exciton Fermi-sea correlations inapplicable. Also, for moderate electron densities the effective exciton-electron coupling constant~\cite{Glazov2020a} $g \sim \sqrt{E_F E_{tr}}$, where $E_F$ is the electron Fermi energy is sufficiently large and becomes comparable with the trion binding energy. As a result, already at the moderate doping, the $2s$ exciton oscillator strength is shuffled to the corresponding charged exciton (attractive Fermi-polaron) resonance.

\newpage

%